\documentclass[preprint2,times,tighten]{aastex6}
\pdfoutput=1 %for arXiv submission

\usepackage{amsmath,amstext}
\usepackage[all]{hypcap} %Figure refs go figures (breaks deluxetables; use \capstartfalse \capstarttrue to fix it)
\usepackage{graphicx}
\usepackage{placeins}
\usepackage{float}
\usepackage{lipsum}
\usepackage{wrapfig}
\usepackage{appendix}

\shorttitle{Hyades Radius Inflation}
\shortauthors{Jaehnig et al.}

\newcommand{\teff}{\ensuremath{T_\mathrm{eff}}}
\newcommand{\teffs}{\ensuremath{T_\mathrm{eff}}s}
\newcommand{\logg}{\ensuremath{\log g}}

\newcommand{\msun}{\ensuremath{M_{\odot}}}

\defcitealias{somers2017}{S2017}
\defcitealias{goldman2013}{G2013}
\defcitealias{douglas2016}{D2016}
\defcitealias{delorme2011}{D2011}

\usepackage[normalem]{ulem}
\usepackage{color}
\newcommand{\kgsins}[1]{\textcolor{black}{#1}}
\newcommand{\kgsdel}[1]{\textcolor{red}{\sout{}}}

\newcommand{\red}[1]{\textcolor{black}{#1}}

\begin{document}

\title{Radius Inflation \kgsins{at Low Rossby Number} in the Hyades Cluster}
\author{Karl Jaehnig\altaffilmark{1,2},
        Garrett\ Somers\altaffilmark{2,3},
        and Keivan G.\ Stassun\altaffilmark{2,1}   
        }
\email{karl.o.jaehnig@vanderbilt.edu}

\begin{center}
\altaffiltext{1}{Department of Physics, Fisk University, Nashville, TN, 37208, USA}
\altaffiltext{2}{Department of Physics and Astronomy, Vanderbilt University, Nashville, TN 37235, USA}
\altaffiltext{3}{Vanderbilt Initiative in Data-intensive Astrophysics Fellow}
\end{center}

\begin{abstract}

Radius inflation continues to be explored as a peculiar occurrence among magnetically active, low-mass stars.
Recently \citet{somers2017} showed that radius inflation among low-mass stars in the young open cluster M45 (Pleiades Cluster) is correlated to the rotation rate\red{:} faster rotators are more inflated.
Here we extend that work to a sample of 68 stars of the older open \red{Hyades} Cluster.
We \red{derive the stars'} spectral energy distributions to \red{measure their} bolometric fluxes. With spectroscopically defined \teff\ and {\it Gaia\/} distances we calculate stellar radii using the Stefan-Boltzmann relation. 
We find numerous stars that exhibit significant (3--4$\sigma$) radius inflation relative to a nominal cluster isochrone.
We compare these results to that of the younger Pleiades and consider radius inflation as a function of open cluster evolution.
We find that unlike the Pleiades, there is not a statistically significant correlation between radius inflation and stellar rotation period.
However, we do find that most inflated stars have (rapid) rotational Rossby numbers of 0.1--0.2, such that the correlation of radius inflation with Rossby number is statistically significant at 99.98\% confidence.
\red{Because the canonical rotation-activity relation of low-mass stars is understood to result from the connection between magnetic activity and surface convection, our results imply that magnetic activity within the convective layers of low-mass stars is what preferentially drives radius inflation}.
\end{abstract}

\maketitle

\section{Introduction}\label{sec:intro}

A consensus is emerging that some low-mass stars ($M \lesssim 1 M_{\odot}$) have larger \red{radii} than expected from standard stellar theory. This so-called ``radius inflation'' effect is typically of order 10-15\% and is correlated with a roughly 5--10\% lower effective temperature than predicted. Radius inflation has been observed in numerous studies using a variety of observational techniques including eclipsing binary analysis \citep[e.g][]{popper97,torres02,lopez-morales05}, statistical projected radii \citep[e.g.][]{jackson16,jackson18}, and spectral energy distributions (SEDs) \citep[][S2017 hereafter]{somers2017}. Though the precise mechanism is still debated, strong magnetic activity seems to play a role either though the direct inhibition of convective energy transport, the influence of large starspots on the photospheric pressure and temperature, or a combination of both effects \citep[e.g.][]{mullan01,chabrier07,macdonald10,feiden13,feiden14,jackson14a,jackson14b,somers2014,somers2015b,somers2015a}. This conclusion is based on observed correlations between radius inflation and proxies of magnetic activity such as H$\alpha$ emission, X-rays, and rotation rate \red{ \citep[e.g.][]{lopez-morales05,stassun2012,somers2017}.}

\citetalias{somers2017} investigated whether any of the single K-dwarfs in the young Pleiades open cluster showed evidence of radius inflation. By \red{measuring} the SEDs of the stars to determine their luminosities, measuring their effective temperature using color proxies, and solving the Stefan-Boltzmann equation, \citetalias{somers2017} determined the radii of 80+ Pleiads and compared them to stellar evolution models. \red{They} found that for stars rotating with a period slower than 2~d, corresponding to a Rossby number\footnote{The Rossby number is defined as the ratio of the rotation period to the convective overturn timescale \citep[e.g.][]{noyes1984}.} (R$_N$) greater than $\sim 0.1$ \red{in the mass range they studied ($\sim 0.7-0.9$\msun)}, the models predicted the \teff-radius relation extremely well. However, stars rotating faster than 2~d (R$_N \lesssim 0.1$) were on average $\sim 12$\% larger than predicted. This is an interesting value of R$_N$, as numerous other studies have found that the correlation between rotation rate and magnetic proxies saturates at approximately this value \citep[e.g.][]{wright2011}. This suggests that the radius inflation mechanism may be connected to the as-yet unclear physics of magnetic saturation.
%clearly demonstrating a rotation-radius inflation correlation in this young cluster.
%A Rossby number of approximately 0.1 has been identified in numerous studies as an important transition point -- at larger Rossby values, magnetic proxies (e.g. X-ray emission) decrease with slower rotation, but at lower Rossby values there is little-to-no change in magnetic proxies with increasing rotation. 
Moreover, these inflated Pleaids tended to show higher lithium abundances as expected from stars experiencing radius inflation during their early lives \citep[e.g.][]{somers2014,somers2015b,somers2015a}. 

Following the results of \citetalias{somers2017}, we now wish to explore the evolution of radius inflation \red{with stellar age}. If radius inflation is a consequence of magnetic activity and rapid rotation, it stands to reason that the degree of radius inflation at fixed spectral type should decline with increasing age as stars both spin down and become less magnetically active \citep[e.g.][]{skumanich1972}. If in fact radius inflation \red{exists for stars with Rossby numbers below} $\sim$0.1 as suggested by the results of \citetalias{somers2017}, then the masses of inflated stars \red{should be lower on average in older clusters}, owing to the slower spin down rate of less massive stars. 
%\textcolor{red}{We should address this point if we haven't already. Does the mass/Teff range where stars are inflated change from Pleaides to Hyades? I think the answer is yes but we should make sure to say this.}
As a complement to the Pleiades, this paper will focus on the Hyades.

The structure of the paper is as follows: In section \ref{sec:data} we discuss the data we collected to form a complete sample of the Hyades pre-main sequence stars as well as any relevant data quality measures that were employed. In section \ref{sec:analysis} we detail how we calculated the radius inflation for each star in the Hyades. In section \ref{sec:results} we present the results we found between the derived radii inflation within the Hyades and stellar rotation of the stars. In section \ref{sec:discussion} we discuss the implications of our results with respect to stellar evolution and with regard to previous work in radii inflation. We present a summary of our findings in section \ref{sec:conclusions} 

\section{\kgsins{Data}}\label{sec:data}

\subsection{The Hyades sample}

%\subsubsection{Source Catalogs}\label{ssec:catalogs}
%\par
In order to properly derive effective temperatures for the stars within the Hyades, it is necessary to build a sample of known members and their magnitude measurements in different bands. We start with the Hyades membership list from the \citet[][G2013 hereafter]{goldman2013} catalog. \citetalias{goldman2013} made use of Pan-STARRS \citep{chambers2016} and PPMXL \citep{roser2010} photometry, complimented by photometry from 2MASS \citep{skrutskie2006}, SDSS-III DR8 \citep{aihara2011}, and UKIDSS DR8 \citep{lawrence2007}, in order to construct a complete stellar mass function. The \citetalias{goldman2013} final sample size of Hyades member stars was 774. The \citetalias{goldman2013} catalog did not have rotational periods for its final sample of Hyades stars and so it is necessary to complement the \citetalias{goldman2013} photometry with other catalogues which included rotation measurements. 

\par
To this end we employ two catalogues which have Hyades member stars with measured rotational periods. \citet[D2016 hereafter]{douglas2016} measured rotation periods for 65 Hyades stars using the Kepler \red{spacecraft} during its K2 phase in order to study low-mass star gyrochronology. We cross-matched the \red{sample compiled by} \citetalias{douglas2016} against the 2MASS catalog in order to find complimentary stars within \red{the membership catalog from \citetalias{goldman2013}}. We find that all 65 of the \citetalias{douglas2016} Hyades stars have cross-matches within the \citetalias{goldman2013} catalog. 

\par
\begin{figure}[tb]
%\hspace*{-1.5cm}
% \epsscale{1.2}
% \begin{center}
\epsscale{0.50}
\includegraphics*[width={\columnwidth}]{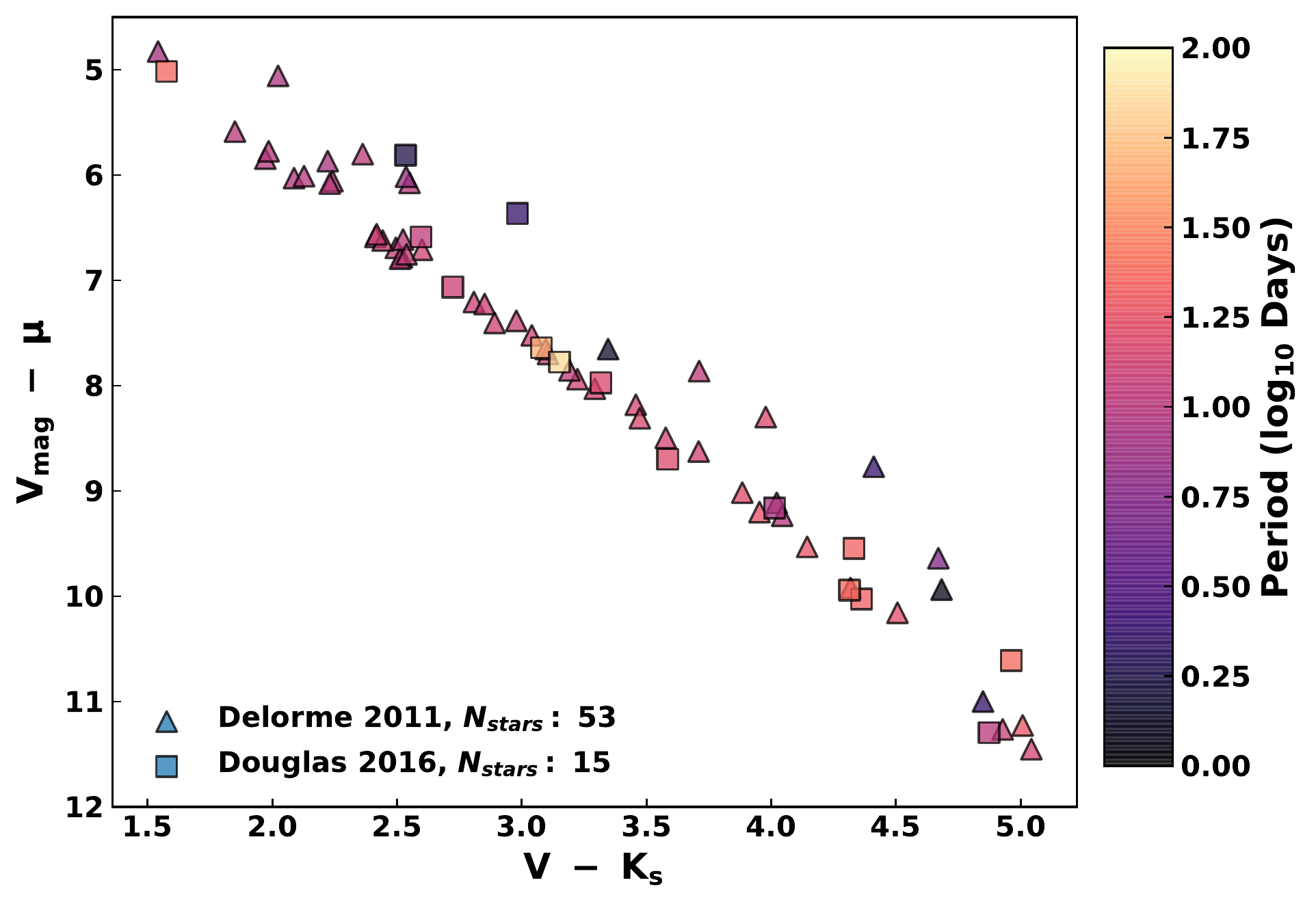}
\parbox{\columnwidth}{\caption{Color-magnitude diagram of the Hyades sample used in this work. Triangles are sources taken from \citet{delorme2011} and squares are sources taken from \citet{douglas2016}. The points are all colored by their period in $\log_{10}$ days.}}
% \end{center}
\label{fig:hyades_source_cat}
\end{figure}

 
\citet[][D2011 hereafter]{delorme2011} also measured rotation periods for 63 Hyades stars in order to also study gyrochronology within the Hyades. \red{We cross matched the catalog} from \citetalias{delorme2011} with \red{the catalog from \citetalias{goldman2013} and find 59 Hyades stars with period\red{-}of\red{-}rotation measurements}. Within this set of 122 stars we check\red{ed} for stars with double observations, finding 13 stars that have observations in both \citetalias{delorme2011} and \citetalias{douglas2016}, thus reducing our initial sample to 109 stars \red{with individual period\red{-}of\red{-}rotation measurements. For Hyades stars with double observations, we employed the period\red{-}of\red{-}rotation measurements from the more recent \citetalias{douglas2016} catalog.} We plot the color-magnitude diagram of this sample in Figure~\ref{fig:hyades_source_cat}.

%\textcolor{red}{Is this CMD with apparent or absolute magnitudes? This should be made clear here. I would support plotting absolute magnitudes because we have DR2 now. Edit(Karl):DONE}

\par
Having compiled a set of Hyades members with measured periods, we move\red{d} to collect photometry in order to calculate the \teff. We quer\red{ied} the SIMBAD database \citep{simbad2000} in order to gather photometry in the  Johnson pass-bands ($UBVR_{J}I_{J}JHK$). \kgsdel{Following the assumptions laid out by \citetalias{somers2017}, we attempt\red{ed} to collect $V$ and $I_{C}$ photometry as this will provide a precise upper limit on the \teff, and therefore a precise lower limit on radius. \red{With a precise lower limit on the radius, it is possible to detect variations on the stellar radius caused by radius inflation.} Unfortunately, \red{as \citetalias{somers2017} found a dearth of measurements in the $I_{C}$ band for the Pleiades, we too found a dearth of $I_{C}$ measurements for the Hyades stars.}} We \red{found} that there is sufficient $V$ and $K_{S}$ 2MASS photometry to calculate \teff\ using the $V-K_{S}$ color. We cross-match\red{ed} 2MASS IDs with CDS to download the $K_{S}$ magnitudes and use the $V$ magnitudes from \citet[]{johnson1955} in order to assemble our sample of colors for the Hyades stars. %\textcolor{red}{Should probably mention the sources for the V-band data. Don't need to go star by star, but something like "We get XX V-band measurements from Soandso (19xx) and XX measurements from Thatguy (20xx). Hopefully there are few enough sources that this isn't onerous, but the photometry gurus deserve their citations.}

%\textcolor{red}{One pedantic note (its a good thing we're at the point for pedantic notes!) -- technically we are not using K-band, but K$_S$-band, which is the particular filter used by 2MASS.}
%, for which we have 109 stars. }. 

\par 
\red{With \red{these} data we assembled our initial sample of 109 Hyades stars with $V-K_{S}$ photometry, and measured rotation periods. In section \ref{ssec:teffective} we go into detail on deriving \teff\ values for our initial sample which pruned our initial sample size of 109 stars to 68 due to constraints on applicable color range for the derived color-\teff\ polynomial fits. In section \ref{ssec:binaries} we detail our search of the previous literature for confirmed spectroscopic/visual binaries within our Hyades sample, finding 23 binaries within our sample of 68 Hyades stars. Our {\bf final sample} of Hyades stars consists of this group of 45 confirmed single stars and 23 identified binaries.} We provide the 2MASS IDs, positions, photometric measurements, and measured rotation period for these 68 Hyades stars in Table~\ref{tab:basic_stellar_data}.

% \begin{figure*}
% %\hspace*{-1.5cm}
% % \epsscale{1.2}
% \begin{center}
% \includegraphics*[width={\eps\columnwidth}]{teff_comparison_vk_vj.pdf}
% \parbox{16cm}{\caption{Plot of \teff\ values derived two different color-temperature polynomial fits. The x-axis is \teff\ values using the Color-Metallicity relation from Mann 2016 using [V-J] colors. The y-axis are the calculated \teff\ values using the \cite{huang2015} fit and [V-K] colors.}} 
% \end{center}
% \label{fig:hyades_tvk_tvj}
% \end{figure*}

\section{\kgsdel{Analysis}\kgsins{Methods}}\label{sec:analysis}

\subsection{\kgsins{Overview}}
% Give a brief summary here of the Pleiades sample from Somers \& Stassun (2017), and simply state that for the analysis in this paper we recalculate the radius inflation of the Pleiades stars using the new approach presented in this paper. 
\par 
The constraints applied to the data that we have just described largely follow the same conventions employed by \citetalias{somers2017} in their analysis of the Pleiades cluster. In the following sections we outline our methods in analyzing our sample of Hyads. We also introduce a novel method for calculating stellar radius inflation that takes into account the relationship between \teff\ and radius (see section~\ref{ssec:radii_inflation}).
\par 
We \red{aim} to compare our results to the Pleiades results from \citetalias{somers2017}, but their method for determining $\Delta R$ did not account for this relationship. Instead, they simply used the isochronal radius at the measured \teff\ of the star as a comparison point. A direct comparison with their values would therefore be biased. Instead, we recompute radius inflation for the Pleiades using the methods we have developed in section~\ref{ssec:radii_inflation}.

\par 

Using the well \red{known} Stefan-Boltzmann law 
\kgsdel{as given below}
\begin{equation}\label{equ:sb_law}
    L=4\pi R^{2}_{*} \sigma_{\rm SB} T_{\rm eff}^{4} ,
\end{equation}
 the radius of a star can be derived if its luminosity and \teff\ are known with acceptable precision. The luminosity can be determined using the equation
\begin{equation}\label{equ:l_flux}
%    f_{bol}=L/4\pi d^{2},
    L = 4\pi d^{2} f_{\rm bol},
\end{equation}
where $f_{\rm bol}$ is the bolometric flux and $d$ is the distance to the star. The \teff\ calculation remains the most crucial task in studying radius inflation, as it is the biggest contributor to the final uncertainty in the calculated radius ($R\propto\ L^{1/2} T^{-2}$). In the subsections that follow, we briefly go over the derivation of \teff\ , the bolometric flux $f_{\rm bol}$, and the stellar radii for our sample of Hyads. \red{The methods used in the derivations of \teff, and $f_{\rm bol}$ are the same used in \citetalias{somers2017}.}   
\kgsdel{but do not go too heavily into detail, as they do not deviate significantly from the analysis of \citealias{somers2017}).}

\subsection{\kgsdel{Deriving} Effective Temperature\kgsdel{s}}\label{ssec:teffective}
% \par 
% The color of a star is the ratio of bluer (shorter-wavelength) light to redder (longer-wavelength) that it emits. The color is frequently used as an analog for the stars \teff\ since the \teff\ delineates the frequency of the black body spectrum peak for the star, in the reference case. There has been a large amount of work done on numerically fitting the color-\teff\ relationship for numerous pass-bands. One of the more prevalent methods is to use fit polynomial functions to well calibrated observations of stars in target pass-bands with robust measurements of \teff\ from spectroscopic observations. Often, the polynomial will have a color range for a particular pass-band in which the fit will be accurate. For our purposes we needed a color-\teff\ fit that was accurate over the V-K color range of our sample of Hyades stars.

% \begin{figure}[tb]
% %\hspace*{-1.5cm}
% % \epsscale{1.2}
% \begin{center}
% \includegraphics[width=\columnwidth]{teff_vk_gaia_rvs_teff_comparison.pdf}
% \parbox{\columnwidth}{\caption{Comparison of our derived \teff [V-K] values with spectroscopically derived \teff values in Gaia DR2 from the RVS instrument.}} 
% \end{center}
% \label{fig:hyades_teff_validation}
% \end{figure}

\par 
In order to calculate \teff\ using the $V-K_{S}$ color we need to account for any possible extinction present in the Hyades. Since the Hyades is both within the local bubble and an older open cluster \citep[650$\pm$70 Myr, from][]{mart2018}, there is very little extinction arising within the cluster or in the intervening ISM. We adopt $E(B-V) = 0.01$ \citep[]{gunn_stryker1983}, a typical value quoted for the Hyades. To convert this value to other colors, we employ the standard reddening law from \cite{cardelli1989} with a selective reddening of R$_{V}$ = 3.1. This gives us an $E(V-K_{S})$ value of 0.027$\pm$0.01. With this we then proceed to use the empirical metallicity dependent calibrations from \cite{huang2015} to calculate \teff\ using $V-K$, adopting [Fe/H] = 0.13$\pm$0.01 \citep{paulson2003}. These color-\teff\ calibrations are valid over a $V-K_{S}$ color range of [0.85, 5.05].

\red{We propogated the uncertainties on our $V-K_{S}$ color, the extinction $E(V-K_{S})$ quoted for the Hyades, and the Hyades metallicity [Fe/H] to get our overall uncertainties on \teff. These individual uncertainties on \teff\ are in the range of 5--34~K. We also had a systematic spectroscopic uncertainty of 60~K calculated from the polynomial fits of \cite{huang2015}. Our overall \teff\ uncertainty ranges over 65--94~K.}
%Gunn & Stryker 1983 derive E(B-V) = 0.01

\subsection{Accounting for Binaries within our Hyades Sample}\label{ssec:binaries}
\red{When it comes to the study of radius inflation in single stars, binaries pose a problem}, as they can produce false indications of radius inflation photometrically. \red{This problem arises from} the additional flux \red{from an unresolved binary system}, causing observations to result in a higher measured magnitude. Typically, photometric binaries are identified because they form their own main sequence above the single star main sequence in the color-magnitude diagram. This effect has already been explored by \citetalias{somers2017} who searched for and excluded binaries within their Pleiades sample. We summarize their conclusions of the effects of binaries on radii inflation below:
\begin{itemize}
    \item Equal mass binaries will increase bolometric flux without significantly affecting the measured \teff, thus leading to a significant signal of radius inflation
    \item For low-mass-ratio binaries, the total bolometric flux is barely affected but the near-IR emission is significantly boosted, leading to a lower inferred \teff\ calculation. This can result in a false signal of radius inflation 
    because the lower inferred temperatures means the star will be compared to the radius predicted for a lower-mass star.
    %as the \teff\ measurement will require a significantly inflated radius to maintain the same bolometric flux.
\end{itemize}

\par 
We proceed\red{ed} to look through our sample of Hyades stars for binary systems that have been confirmed with previous surveys. We focus\red{ed} primarily on finding confirmation either through spectroscopic surveys, which involve measuring the radial velocity and possible calculation of the orbital elements, or optical surveys\red{,} where the binary is resolved and its motions can be calculated. 

\par 
As the Hyades is among the most observed open clusters, it \red{was} not difficult to find several surveys that considered both cluster membership and binarity. For our sample of Hyades stars we consult\red{ed} \cite{kopytova2016} for an initial list of their identified single stars, as well as the works they used to exclude stars that were in binary systems. We construct\red{ed} a list of confirmed multiple systems using the catalogs of \cite{mermilliod2009}, \cite{patience1998}, and \cite{duchene2013}. \red{With this compiled list of catalogs } we found that 45 stars within our sample of 68 Hyades stars had been previously identified as single stars. 

\par 
For the other 23 stars within our sample we queried Vizier for any catalogues in which the binary status had been confirmed \red{visually or spectroscopically.} We found that 20 of our 23 Hyades stars are confirmed binaries, with observations coming from the Tycho Double Star Catalog\citep{fabricius2002}, the Washington Double Star Catalog\citep[]{mason2001, douglas2014}. We \red{could not find} confirmation of binarity, or single star status for three stars and \red{decided} to exclude them from our analysis. The final binary status for our final sample can be found in the last column of Table~\ref{tab:basic_stellar_data}, where
\begin{itemize}
    \item No\ =\ Confirmed Single Star
    \item Yes\ =\ Confirmed Binary System
    \item Yes?\ =\ Unconfirmed Binary/Single Star. 
\end{itemize}

\subsection{\kgsdel{Calculating} Bolometric Flux}\label{ssec:bolometricflux}
\par
In order to calculate the total bolometric flux of a star, $f_{\rm bol}$, we \red{interpolated} the observed spectral energy distribution (SED) of each Hyades star with a standard stellar atmosphere model from \red{the model grid of} \cite{kurucz2013}. This is the same SED \red{interpolation} procedure employed in \cite{stassun_torres_2016} and was used in \citetalias{somers2017}. \kgsdel{The stellar atmosphere models used here assumed main-sequence surface gravity \logg\ and solar metallicity as was done in \citetalias{somers2017}.} Note that, as described by \citet{stassun_torres_2016}, the $f_{\rm bol}$ determined via this interpolation procedure is virtually independent of \teff. 

%Each stellar atmosphere model was initiated at the 
\red{We adopted the Hyades metallicity and the} \teff\ derived in Section~\ref{ssec:teffective}. The multiple photometric observations were collected by querying {\it Vizier} and cover a wide wavelength spectrum from the far-ultraviolet at $\sim 0.15\mu m$ to the far-infrared at $\sim 22\mu m$. The complete list of surveys queried can be found in Section 2.3.1 of \citetalias{somers2017}. \red{With these collected photometric observations spanning a wide wavelength spectrum it was then possible to construct an SED to calculate the total $f_{\rm bol}$.} The resulting SEDs have a goodness-of-fit, $\chi_{\nu}^{2}$, 
%value associated with them that was calculated from the case of 
\red{determined by comparing the observed fluxes to the passband-integrated model fluxes, and where the only free parameters in the fit are the reddening (limited to the allowable range determined above) and the overall flux normalization}. 
%\red{[KEIVAN, COULD YOU CLARIFY HERE HOW YOU GET CHISQ2 FROM THE FIT OF AN INTERPOLATED CURVE?]}. 
These values are listed in Table \ref{tab:derived_stellar_data}, along with the calculated bolometric flux. 

\subsection{\kgsdel{Calculating Observed Angular} \kgsins{Stellar} Radius}\label{ssec:angradii}
\par 
The angular radius can be derived using a rearranged version of the Stefan-Boltzmann law, accomplished by replacing the luminosity in equation\ \ref{equ:sb_law} with the luminosity-flux relation given in equation\ \ref{equ:l_flux}. The resulting equation can be rearranged such that we get
\begin{equation}\label{equ:ang_radii_equ}
    \Theta = f_{\rm bol}^{0.5}\ \sigma_{\rm SB}^{-0.5}\ \teff^{-2}
\end{equation}
where $\Theta$ is the angular radius, given by $R_{*}/d$. As we have already calculated the total bolometric flux, as well as the \teff, we can simply calculate the angular radius for our Hyades Sample by inserting these values into equation\ \ref{equ:ang_radii_equ}. We list the resulting angular radii in units of milli-arcseconds in Table \ref{tab:derived_stellar_data}. 

\kgsdel{We plot the angular radius $\Theta$ against the angular radius uncertainty $\sigma_{\Theta}$ in Figure \ref{fig:hyades_ang_radii}, where the points are sized by their bolometric flux, and colored by the calculated $\chi^{2}_{\nu}$ of the SED interpolation. The stars with red crosses are the stars identified as binaries within our sample.} Our calculated angular radii for our identified single stars have a range of [35-100]\ $\mu as$ with an average error on the calculated angular radius of $\sim$2.5$\ \mu as$. Overall, the small magnitude of the angular radius error, along with the relatively small values of $\chi^{2}_{\nu}$ indicate that the SED fits were accurate.

% \begin{figure}
% \hspace*{1.5cm}
%  \epsscale{1.2}
%  \begin{center}
% \plotone{angular_radius_plot_figure_3.pdf}
% \parbox{\columnwidth}{\caption{Plot of the angular radius uncertainty $\sigma_{\Theta}$ against the angular radius $\Theta$. Both the x-axis and the y-axis are in units of micro-arcseconds. The points are colored according to the calculated $\chi_{\nu}^{2}$ value of the SED fit. The size of the points are proportional to their $10^{10} \cdot F_{bol}$. The points with red crosses are identified binary stars.}} 
%  \end{center}
% \label{fig:hyades_ang_radii}
% \end{figure}

%\subsubsection{Calculating Observed Stellar Radii}\label{sssec:stellar_radius}
To get the stellar radius from the angular radius, we \red{multiplied} the angular radius $\Theta$ of a star by its distance from us. The distance to the Hyades has been found to be about 46.75$ \pm\ $0.46~pc, using TGAS data from {\it Gaia\/} Data Release 1 \citep{gaia_collab_2017}. \citetalias{somers2017} employed a single star cluster distance measurement for the Pleiades to convert their angular radii derivations to stellar radii, with an uncertainty added to the distance to reflect the depth of the Pleiades cluster. 

\par 
With the recent release of {\it Gaia\/} Data \red{R}elease 2 \citep{gaia_dr2_paper}, we now have individual parallax measurements with very high precision ($\sigma_{\pi} \sim 10^{-4}$~arcsec) for our entire Hyades sample. \red{Instead of simply inverting the Gaia DR2 parallaxes to derive distances, we employed the distances calculated by \cite{bailor_jones_2018} using Bayesian Inference for our individual Hyades stars.} We then derived the observed stellar radii by multiplying our individual angular radii values by the associated individual distances we have for the Hyades stars. We list these stellar radii values in units of $R_{\odot}$ in Table~\ref{tab:derived_stellar_data}.

\par
\begin{figure}
%\hspace*{-1.5cm}
% \epsscale{1.2}
\begin{center}
\includegraphics[width=\columnwidth]{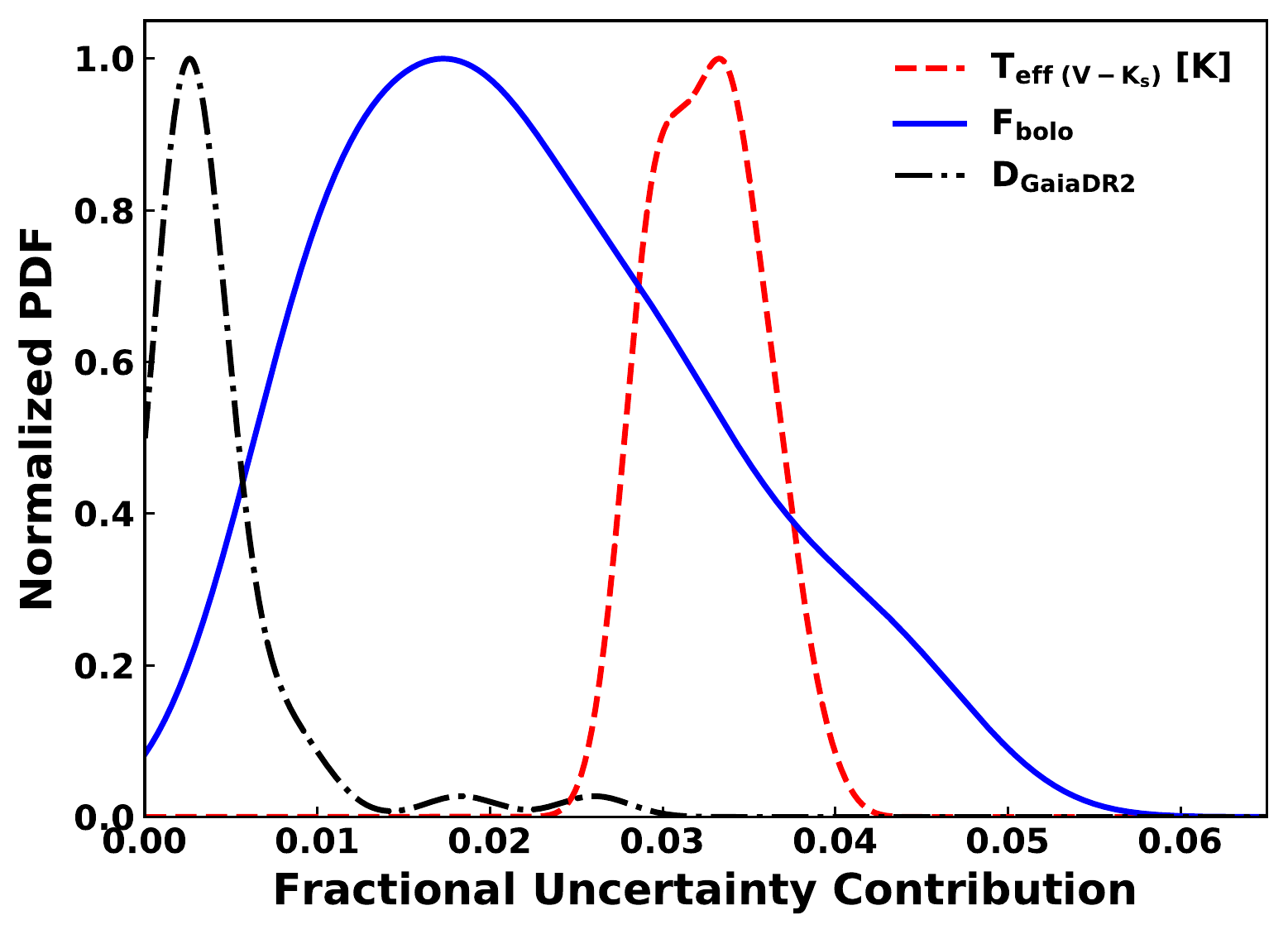}
\parbox{\columnwidth}{\caption{\red{Normalized probability distribution functions of the fractional uncertainties that $f_{\rm bol}$, \teff, and Distances from {\it Gaia\/} DR2 that contribute to the error on the calculated stellar radius. These PDFs have been generated from Kernel Density estimators that were fitted to the distributions for the final sample of single stars. Here we see that the typical contributions to the uncertainty on $R_*$ are $\lesssim 1\%$, $\lesssim 4\%$, and $\lesssim 3\%$, for distance, \teff, and $f_{\rm bol}$, respectively.}}} 
\end{center}
\label{fig:hyades_frac_errors}
\end{figure}

\red{The stellar radii are calculated using three other parmeters, each with their associated uncertainties.}  These parameters are the derived \teff, the total bolometric flux, and the distances for each star. \red{We fitted a gaussian kernel density estimator to the distributions of fractional errors for the three aforementioned parameters and plot the resulting probability density functions in }Figure \ref{fig:hyades_frac_errors}. \red{From Figure \ref{fig:hyades_frac_errors} it is apparent that the error in $\Theta$ and by extension, $\Delta R_{*}$, is dominated by the error on \teff, though can be dominated by the error on $f_{\rm bol}$ in some cases; the error contribution from the {\it Gaia\/} parallax is negligible in all cases}. 
%However, the bulk of the uncertainties on the bolometric flux are under 10$\%$ 
\par
%The fractional errors on the \teff\ derivations, as well as the distances from Gaia DR2 are both well below $5\%$, with the bulk of the \teff\ uncertainties being under $2\%$ and the Gaia DR2 distance errors being under $1\%$.
% It is easier to compare the fractional error, as it scales the two parameters. For the fractional distance errors, we find an average of 0.66 \% and 0.4 \% errors on the total sample, and the single star sample of Hyads, respectively. For the fractional angular radii error we find an average of 3.6 \% and 3.5 \% errors on the total sample, and single star sample of Hyads, respectively.  

%\textcolor{red}{It would be useful here to compare the average error of the angular radius to the average distance error. This should convince readers that distance errors are essentially meaningless for our final errors, and that all uncertainty comes from our Teff and SED fitting.}

\begin{figure}
%\hspace*{-1.5cm}
% \epsscale{.5}
\begin{center}
\includegraphics[width=\columnwidth]{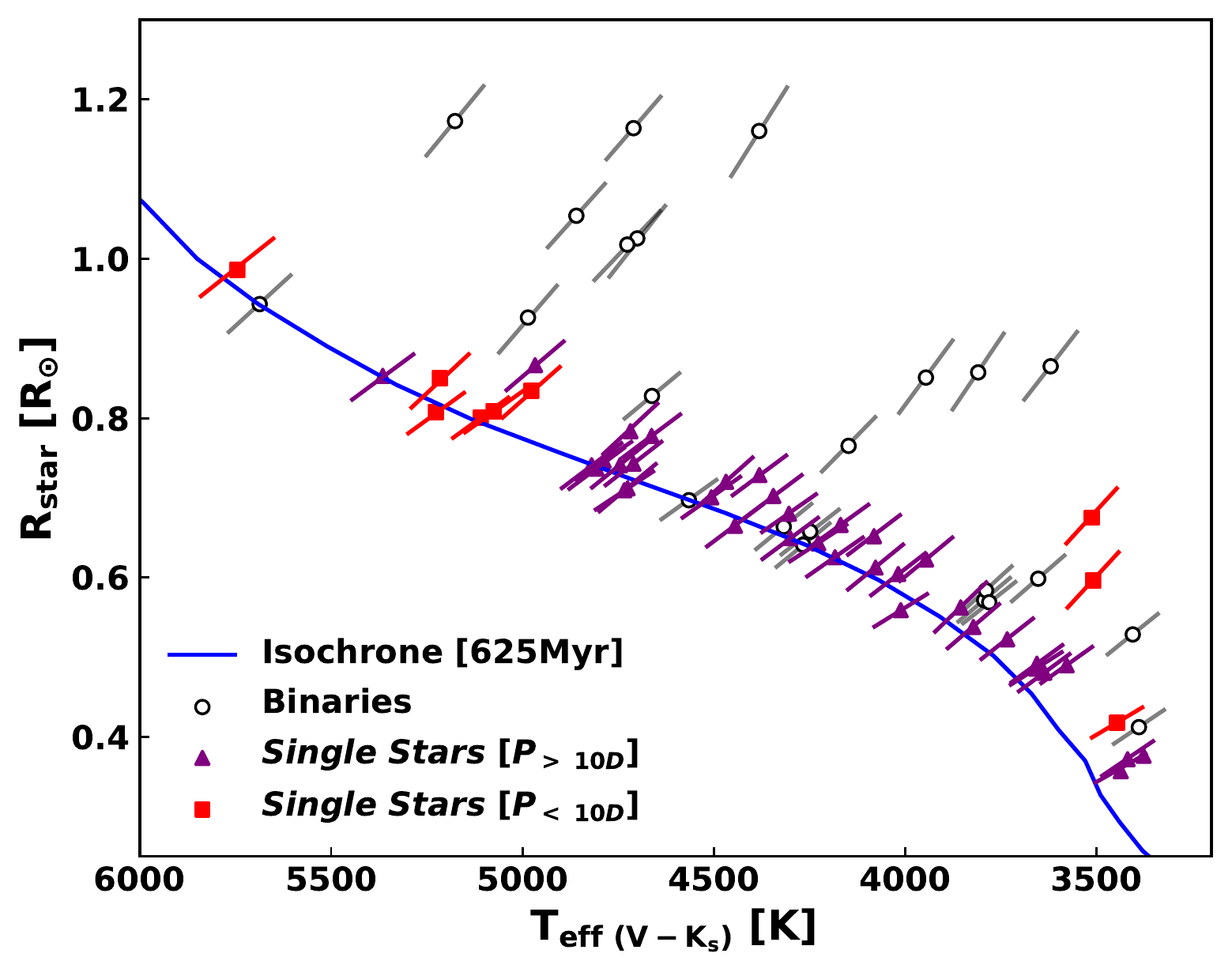}
\parbox{\columnwidth}{\caption{Calculated stellar radii for the Hyades stars against their calculated \teff\ values. The diagonal error bars are the correlated errors between the stellar radius and the \teff. The red squares are stars with measured rotation periods less than 10 day. The purple triangles are stars with measured rotation periods greater than 10 days. The empty gray circles are the identified binaries within the sample. The solid blue line is the isochrone with a cluster age of 625 Myr.} }
\end{center}
\label{fig:hyades_isochrone_wo_calibration}
\end{figure}

\subsection{Calculating Radius Inflation}\label{ssec:radii_inflation}
\par 

In order to determine if a star has a larger radius than expected, we compared our derived values to stellar isochrones from the literature. 
Isochrones are frequently used to predict theoretical properties of stellar populations under the assumption that they are all the same age and composition.
%These isochrones are typically constructed using arguments such as metallicity, and age to derive stellar parameters such as stellar radius, \teff, luminosity, and surface gravity.
In modeling the Hyades, we decided to use isochrones generated by \cite{somers2014} with a cluster age of 625 Myr and a metallicity of [Fe/H] = 0.13. We plot this isochrone alongside our Hyades sample in Figure \ref{fig:hyades_isochrone_wo_calibration}. 

\par
The method employed in calculating $R_{iso}$ for the Pleiades in \citetalias{somers2017} (see their section 3.3) is summarized here. Calculating the amount of radius inflation occurring with a particular star requires the measured stellar radius and the expected stellar radius based on \teff\ from a representative isochrone. To calculate the isochronal stellar radius, $R_{iso}$, a spline was fitted to the isochrone relating radius and \teff. Then $R_{iso}$ is calculated for a particular star by using the fitted spline to calculate the radius at its calculated \teff. 

\par 
We introduced a new method for calculating the expected stellar radius $R_{iso}$ for a star at a measured \teff. \kgsdel{Observational results show that radius inflation of $\sim 10$\% is accompanied by a decline in \teff\ of approximately half this percentage \citep[e.g.][]{stassun2012}. It is therefore important to account for this effect when establishing the model radius we compare to the observed radius.} \red{As we assume that the luminosity will remain the same whether or not a star is undergoing radius inflation, we took advantage of this by calculating the luminosity, as given by equation \ref{equ:sb_law} of our Hyades stars using $R_{*}$ and \teff.}

\par
\red{We fit a univariate spline to the isochrone radius and luminosity from the 625 Myr isochrone. With this spline, we can then calculate $R_{iso}$ for each of our Hyades stars.}
We can now proceed in calculating the radius inflation that might be taking place within the Hyades. We employ the same equation from \citetalias{somers2017}, which is written as:
\begin{equation}\label{equ:delta_radii}
    \Delta R_{*} = \frac{R_{*} - R_{iso}}{R_{iso}}
\end{equation}
Where $R_{*}$ is the measured stellar radius we derived in section \ref{ssec:angradii} and $R_{iso}$ is the radius we calculated using isochrones and the calibrated \teff. Thus the $\Delta R_{*}$ value can be more simply stated as the fractional height above the fitted isochrone.

\kgsdel{Consequently, we employed a method which finds $R_{iso}$ for a particular star by moving within radius-\teff\ parameter space bi-linearly from the observed stellar radius to $R_{iso}$ following an equation relating fractional changes between \teff and radius. This relation can be written as follows:
% \vspace{-1.5cm}
\begin{equation}\label{equ:frac_r_teff}
    \frac{dR}{R} = - \frac{2d\teff}{\teff}
\end{equation}
% \vspace{-1cm}

Equation \ref{equ:frac_r_teff} can be derived from equation \ref{equ:sb_law} with the imposed constraint that the change in luminosity, dL, be zero. The full derivation can be found within the appendix of this work. 

\par 
Physically, equation \ref{equ:frac_r_teff} can be said to represent how a star increases(decreases) fractionally in radius for fractional decreases(increases) in its \teff. If a star is undergoing radius inflation, the \teff\ we observe from the V\ -\ K$_{s}$ color is lower than it would be were it not inflated and thus the isochronal radius at the measured Teff is not the appropriate comparison value. Instead, $R_{iso}$ must be found by finding the \teff\ at which the star in question would lie on the isochrone {\it were it not inflated}. This relation is approximately linear for small enough changes in either radius or \teff. 

\par For each Hyades star, we found whether the star's observed stellar radius lies above(below) the isochrone. Then we incremented the \teff\ in a step-wise fashion down(up) and recalculated the final radius after the change in \teff\ using equation \ref{equ:frac_r_teff}. In order to maintain the linearity of the relation, we proceeded in steps of $1\times 10^{-2}$ K. This process is iterated over until the calculated \teff\ is close enough to the \teff\ from the isochrone line, which we take as the difference between the two \teff\ values and impose a difference no greater than $1\times10^{-3}$ K. 

\par 
To visually illustrate how this procedure works, we plot a few selected stars in Figure \ref{fig:hyades_isochrone_process}(right panel) where the solid gray lines represent the calculation of $R_{iso}$ from directly taking the same \teff\ as we calculated in section \ref{ssec:teffective}. \textbf{The red step lines are not representative of the actual step-size used to get the \teff\ without radius inflation.} As the step sizes we employed were 0.01 K, the overall step-wise behavior would look like a straight line. We also plot our entire sample and the newly calculated \teff\ values in Figure \ref{fig:hyades_isochrone_process}(left plot), where the gray dot-dashed lines intersect the isochrone at the \teff\ at which the stars would be, if there was no radius inflation. The symbol colors and shapes of the left plot are the same as in Figure \ref{fig:hyades_isochrone_wo_calibration}. }

% \begin{figure}
% % \hspace*{1.5cm}
% \epsscale{0.5}
%  \begin{center}
% \includegraphics[width=\columnwidth]{hyades_radius_teff_iso_w_calibrations_multi.pdf}
% \parbox{\columnwidth}{\caption{Top panel: Scatter plot of the slow rotating Hyades stars (yellow crosses), the fast rotating Hyades stars (red crosses), and the Hyades identified as binaries (purple crosses). Also plotted is the isochrone generated by \cite{somers&pinsonneault2014}. The gray dash-dotted lines show the difference in \teff\ and radius between any one Hyades star and its actual location on the isochrone. Bottom panel: Zoom-in of the top panel, in which we try to highlight the overall behavior of the DelR-DelT method with some example Hyades stars.}}
%  \end{center}
% \label{fig:hyades_isochrone_process}
% \end{figure}

\section{Results}\label{sec:results}
\par
\begin{figure}
%\hspace*{-1.5cm}
% \epsscale{1.2}
\begin{center}
\includegraphics[width=\columnwidth]{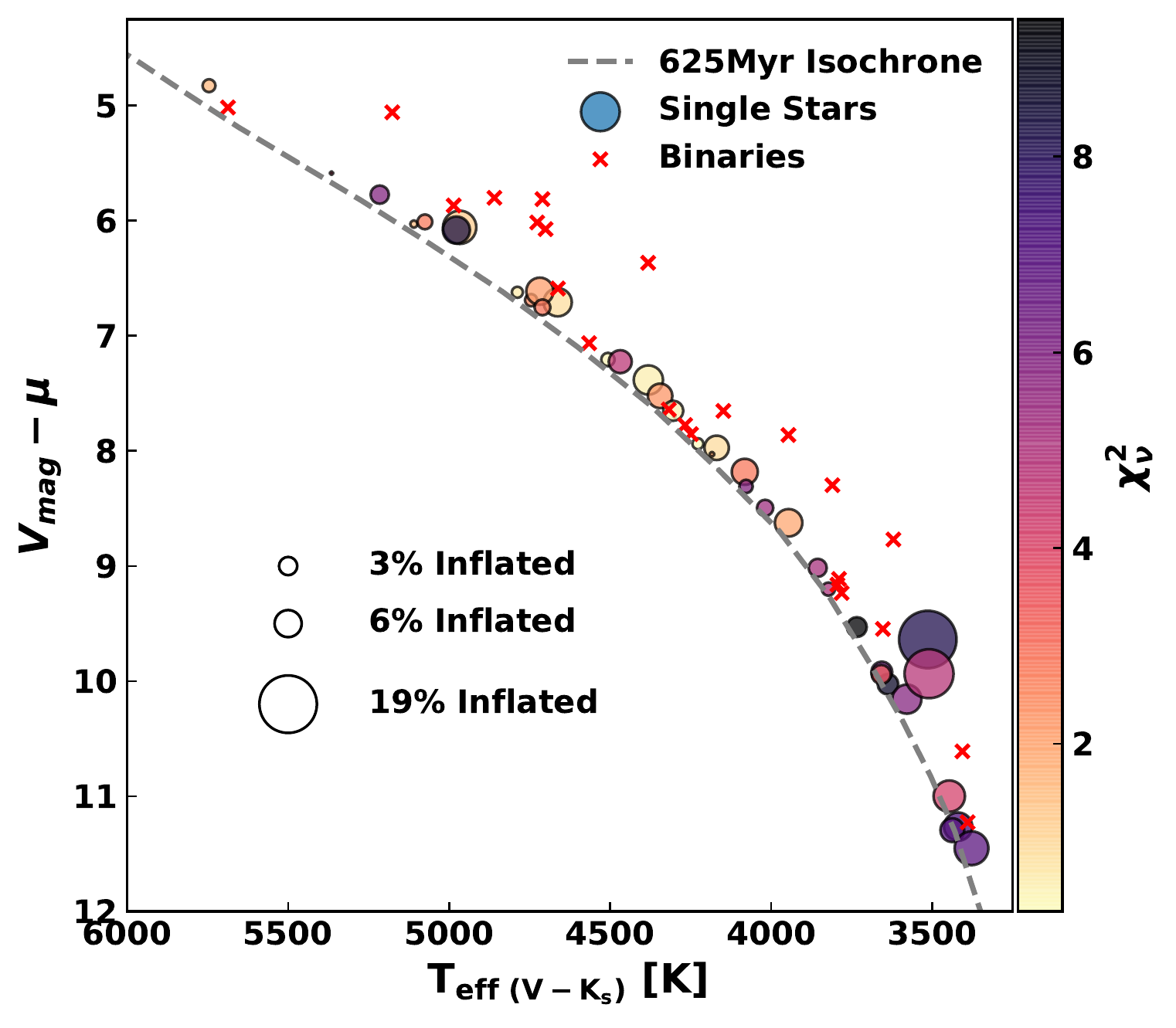}
\parbox{\columnwidth}{\caption{Temperature-magnitude diagram of the Hyades sample. The points are sized proportionally to their calculated $\Delta R_{\star}/ R_{\star}$ and colored according to the $\chi^{2}$ values for their SED fits. The gray dashed line is a 625Myr isochrone from an atmospheric model from Somers et al.\ 2019 (in preparation). }} 
\end{center}
\label{fig:hyades_cmd_delr}
\end{figure}

 
We list the calculated overall radius inflation, $\Delta R$, for each star within our final sample in table \ref{tab:derived_stellar_data}. We also re-plot the color-magnitude diagram of the Hyades within Figure \ref{fig:hyades_cmd_delr} to include the radius inflation information. confirmed single stars within our sample are colored by the $\chi^{2}_{\nu}$ values of the fitted SED and sized according to the calculated radius of inflation, using equation \ref{equ:delta_radii}. Confirmed binary stars within the Hyades are plotted as red x's and not sized according to their degree of radii inflation. \citetalias{somers2017} showed how their properties affect the calculation of radius inflation, causing false signals of radius inflation. Within Figure \ref{fig:hyades_cmd_delr} we see the presence of the binary main-sequence, as well as the location of two inflated single stars within this binary main-sequence. Single stars that have undergone radius inflation are thought to exist within what is considered the binary main-sequence and are incorrectly classified as photometric binaries. We will comment on this further with respect to our results \kgsdel{with}in section~\ref{sec:discussion}.

\subsection{\kgsins{Relationship to Rotation Period}}
\par 
As mentioned in $\S$\ref{sec:intro}, radius inflation may be connected to rapid rotation. We therefore considered the connection between radius inflation and stellar rotation, as was considered within the younger Pleiades open cluster in \citetalias{somers2017}, by looking at the observed rotation period of our Hyades sample and the radius inflation. In Figure \ref{fig:period_rossby_twoplot} (left panel), we plot the observed rotation periods of our single stars against their calculated percentage of radii inflation. Overall we found that our Hyades sample periods are clustered around 10--15 days. 

\par 
We binned the Hyades stars into two bin: the slow rotating stars (period $>$ 10 days) and the fast rotating stars (period $<$ 10 days). Within these two bins we found the average percentage of radius inflation and plot the two average values as the orange squares within Figure \ref{fig:period_rossby_twoplot} (left panel). 
\par 
We explored the degree of monotonicity between radius inflation and rotation period by calculating the Spearman's Rank correlation coefficient, denoted by $\rho_{spearman}$ and the Kendall Tau correlation coefficient, denoted by $\tau_{kendall}$. These coefficients describe the degree of correlation between two variables, where a value of 0 is no correlation, values $>$0 are positive correlations, and values $<0$ are negative correlations. In addition to the actual correlation coefficient, we calculated their p-values, to test against the null-hypothesis that these correlations could be due to random chance. 

\par 
For period and $\Delta R$, we calculate a value of $\rho_{spearman} = -0.032$ and $\tau_{kendall} = -0.022$, with p-values of 0.832 and 0.829, respectively. This suggests that there is little to no correlation between period and radius inflation within this sample.

\par
This result suggests that the relation between rotation and radius inflation found by \citetalias{somers2017} becomes undetectable via observed rotation periods or disappears as the cluster ages and the star's magnetic fields become less active due to age related spin-down. To explore the likelihood of the second possibility, we considered also the Rossby number \citep{noyes1984} of the stars within the Hyades sample. 

\subsection{\kgsins{Relationship to Rossby Nubmer}}
\par 
The Rossby number ($R_{\rm N})$ is the ratio of the rotation period of a star over the convective zone overturn time, denoted by $\tau_{cz}$, which is a timescale to describe how long it takes convective motions to traverse the convective envelope. \citet[][]{noyes1984} argued that the Rossby number is the preferential metric to study magnetic activity because it appears to correlate with magnetic proxies such as Ca~II emission more strongly that just rotation itself. 

\par 
In order to calculate the Rossby number for our Hyades sample, we employed the empirically calculated relationship between \teff\ and the convective overturn time, $\tau_{cz}$ from \cite{wright2011}. We fitted a cubic spline to this empirical relationship and calculated the expected convective overturn rate for our Hyades sample, using their individual \teff\ from section \ref{ssec:teffective}. We calculated the Rossby number for our Hyades sample by dividing the rotation period of our stars by their calculated $\tau_{cz}$ value. 

\par
\begin{figure}
%\hspace*{-1.5cm}
% \epsscale{1.2}
\begin{center}
\includegraphics[width=\columnwidth]{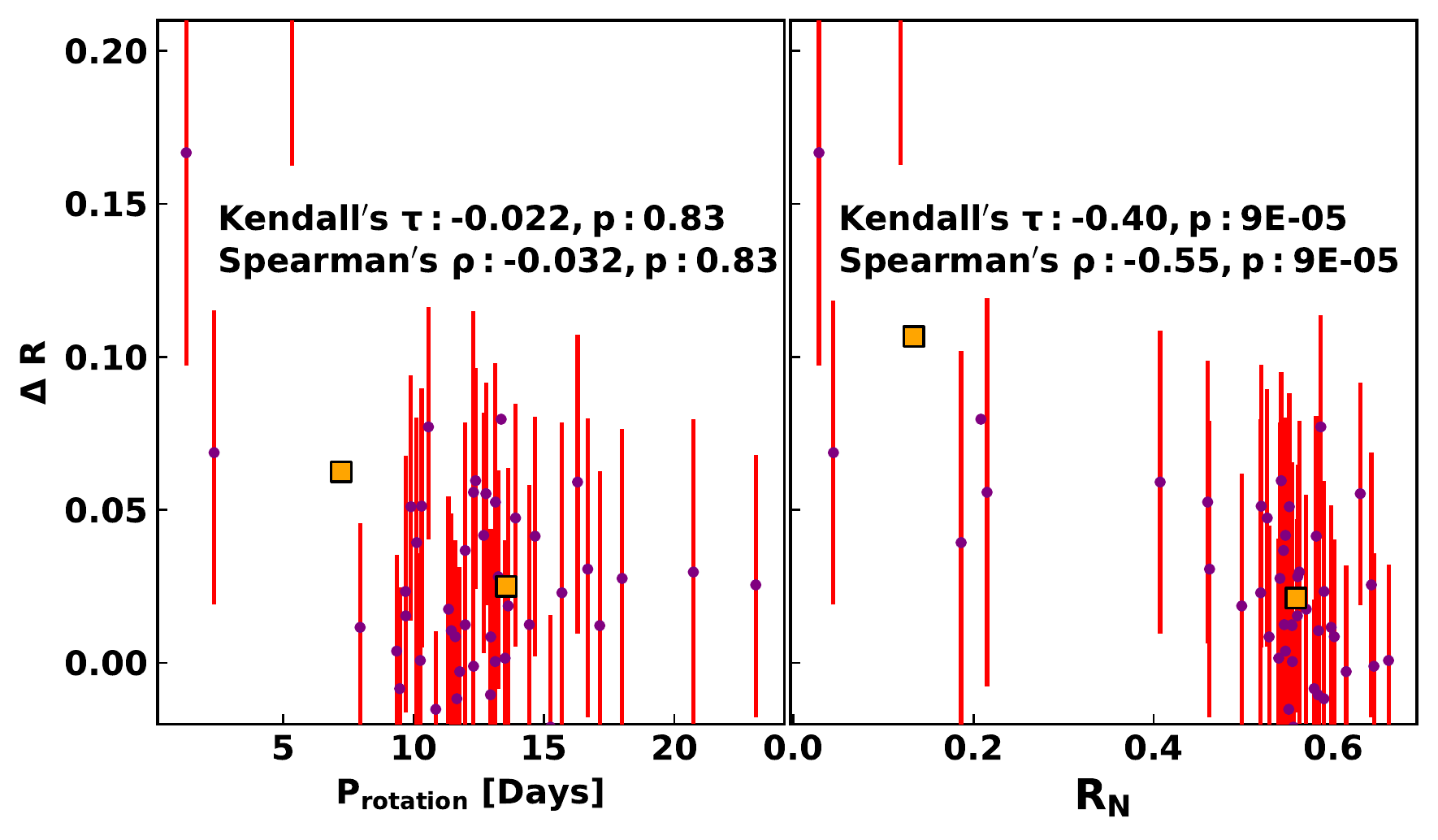}
\parbox{\columnwidth}{\caption{Left Panel: Measured rotation period of the single stars in the Hyades final sample against the calculated percentage of radius inflation from Section\ \ref{ssec:radii_inflation}. The purple points are the confirmed single stars. The orange squares are the average of the sample, binned into slow rotators ($P_{\rm rot}$ $>$ 10 d) and fast rotators ($P_{\rm rot}$ $<$ 10 d). Right Panel: Same as left panel but with the calculated Rossby number ($R_{\rm N}$) on the x-axis. The orange squares are the average of the sample, binned into slowly convective ($R_{\rm N}>$ 0.25) and rapidly convective ($R_{\rm N}<$ 0.25).}} 
\end{center}
\label{fig:period_rossby_twoplot}
\end{figure}

 
We plot the Hyades Rossby numbers against the calculated percentage of radius inflation in Figure \ref{fig:period_rossby_twoplot} (right panel). We bin this sample by either being slowly convective (R$_{0}$ $>$ .25) or rapidly convective (R$_{0}$ $<$ .25)
 We see there there is a greater spread in the distribution of Rossby numbers within the Hyades sample than with period, indicative of the fact that most of the rapidly spinning stars are lower mass. We calculated the $\rho_{spearman}$ and $\tau_{kendall}$ coefficients and attain values of $-0.549$ and $-0.404$, respectively. The associated p-value are both $\sim$0.0001. This indicates that there is significantly non-random negative trend between the Rossby number and the percentage of radius inflation.

\begin{figure}
%\hspace*{-1.5cm}
% \epsscale{1.2}
\begin{center}
\includegraphics[width=\columnwidth]{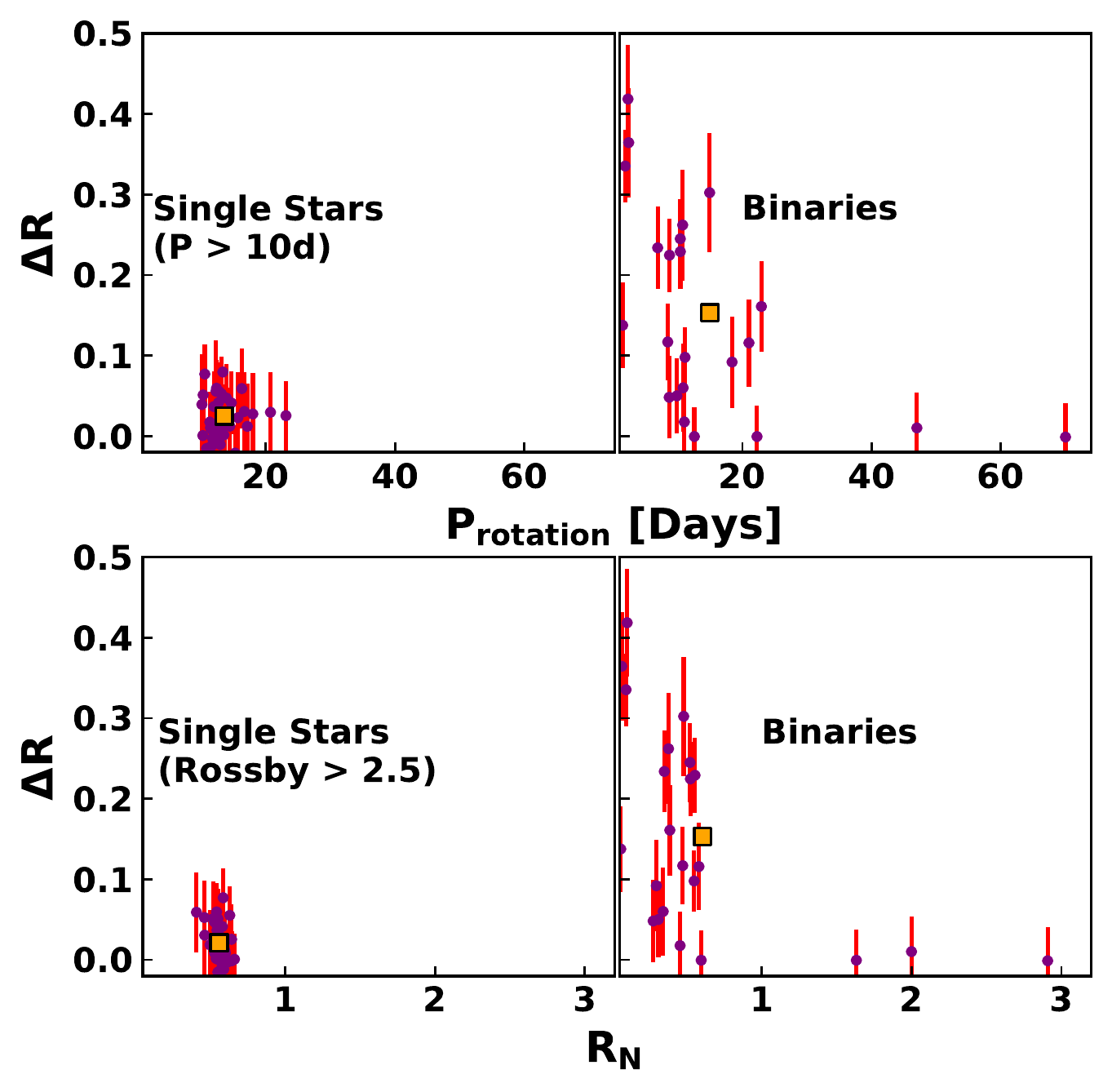}
\parbox{\columnwidth}{\caption{Top Row: $\Delta R$ versus rotation period for slowly rotating single Hyades stars (left panel) and the confirmed binary stars (right panel). Bottom Row: $\Delta R$ versus Rossby number ($R_{\rm N}$) for ``slowly convective" ($R_{\rm N}$ $>$ 0.25) single stars (left panel) and the confirmed binary stars (right panel). The orange square in each panel is the average of the plotted sample. }} 
\end{center}
\label{fig:period_rossby_fourplot}
\end{figure}

\begin{figure}
%\hspace*{-1.5cm}
% \epsscale{1.2}
\begin{center}
\includegraphics[width=\columnwidth]{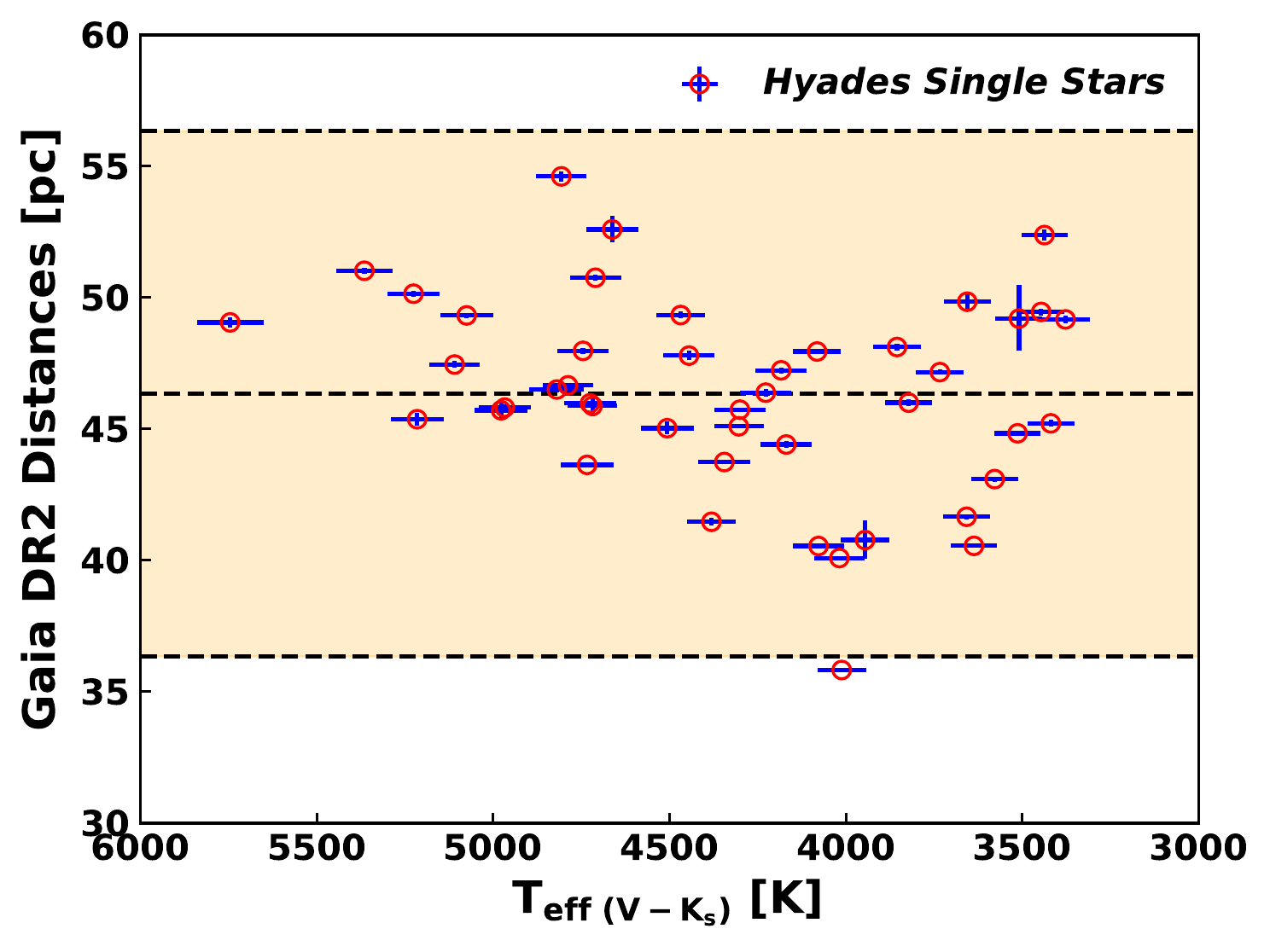}
\parbox{\columnwidth}{\caption{Inferred distances from {\it Gaia\/} DR2 parallaxes \citep[see][]{bailor_jones_2018} versus \teff. Error bars are included for both the distances and \teff. The gray solid line at 46.75~pc is the typically quoted distance to the center of the Hyades cluster. The two black horizontal dashed lines delineate the tidal radius of 10~pc for the Hyades; this region is shaded.}} 
\end{center}
\label{fig:hyades_teff_gaia_dist}
\end{figure}

\begin{figure*}
%\hspace*{-1.5cm}
% \epsscale{1.2}
\hspace*{-0.5cm}
\includegraphics[scale=1]{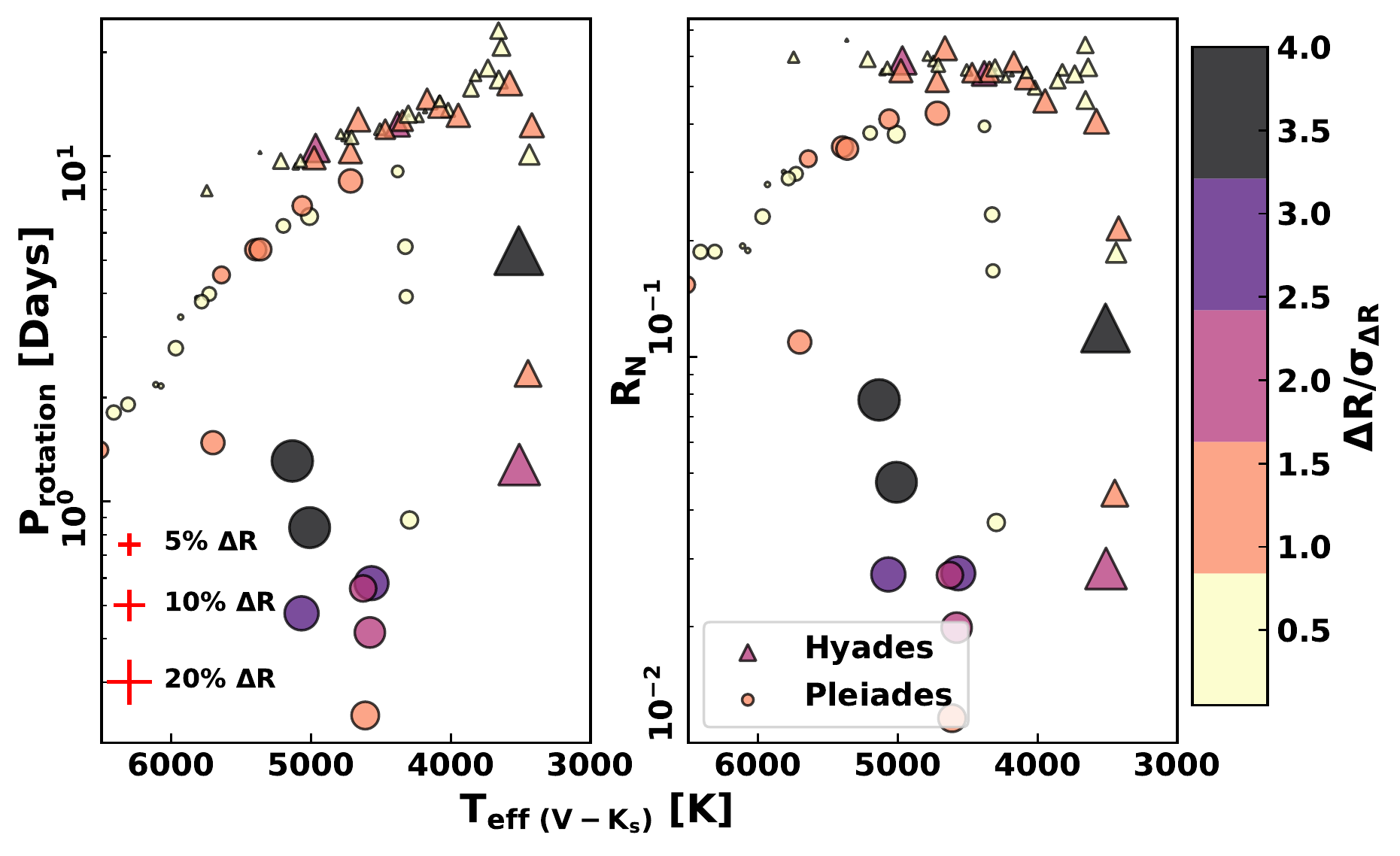}
\begin{center}
\parbox{16cm}{\caption{Left plot: Plot of the rotation period and \teff\ for the Hyades single stars (filled-triangles) and the Pleiades single stars (filled-circles). Right plot: Same as the left plot but now plotting Rossby number($R_{\rm N}$) and \teff.  All points within the two panels are sized proportionally to their calculated value of radius inflation ($\Delta R$). The three red crosses in the right panel serve as reference baselines. All points in the two panels are colored according to the sigma significance of the $\Delta R$. }}
\end{center}
\label{fig:period_rossby_twoplot2}
\end{figure*}

% \begin{figure*}
% %\hspace*{-1.5cm}
% % \epsscale{1.2}
% \begin{center}
% \includegraphics[scale=0.85]{hyades_rossby_num_delR_plot.pdf}
% \parbox{16cm}{\caption{Plot of the $\Delta R_{\star}/ R_{\star}$ values for the Hyades sample against their calculated Rossby number. The purple points are the confirmed single stars within our final sample. The orange squares are the average of the final sample that has been binned by slow convective overturn (Rossby $>$ 0.3) and fast convective overturn (Rossby $<$ 0.3). The blue trend line is }} 
% \end{center}
% \label{fig:delr_rossby}
% \end{figure*}

\section{discussion}\label{sec:discussion}
\par
 We have employed several measures to ensure that our data \red{are} free of spurious values and that the relationship we have found between radius inflation and rotational period is not random. In the following section we discuss possible sources of error and the impact of these sources of error on our findings within the Hyades cluster. The rest of the section will be dedicated to a discussion of our findings in the Hyades with respect to the findings of \citetalias{somers2017} in the Pleiades. We will also discuss radius inflation as a function of age using these two clusters.

\subsection{Potential Sources of Errors}
\subsubsection{Binaries}\label{sssec:binaries}
%We consider our Hyades sample by taking a look at the distributions of radius inflation in our slowly rotating single star calculations against the known binaries. This serves as a benchmark against which we can be assured that the radius inflation we find within our faster rotating stars is not due to something other than stellar rotation. As S2017 showed before, there is a significant contribution to the bolometric flux from a binary system, causing higher angular radius measurements, and therefore increased radius inflation calculations. S2017 compared the periods of their slow rotating Pleiads, and their binaries against the measured radius inflation. They found significantly disparate distributions between the two to confidently state that they had removed binaries effectively from their final single-star data. 
As discussed above, binaries are a potential source of false-positive signals of radius inflation \red{since} the bolometric flux enhancement from the presence of a secondary would appear as higher luminosity -- thus larger radius at fixed temperature -- in our analysis. While it is nearly impossible to prove that a star does not have a binary companion, we can assess how efficiently we have excluded binaries by comparing the derived radii of presumed-single and confirmed-binary stars rotating slower than $R_N \sim 0.1$. As these stars are not expected to be inflated by magnetic activity, stars with anomalously large derived radii can be confidently marked as binaries. \citetalias{somers2017} did precisely this, finding significantly disparate distributions between the two samples. This provided confidence that most, if not all, binaries had been excluded from the full sample.
%\textcolor{red}{I took another stab at this paragraph. Let me know if you like it.}

\par 
We repeat here this same exercise. In the top half of Fig.\ref{fig:period_rossby_fourplot}, we plot the periods of single stars with a rotation period greater than 10 days and binaries against their measured radius inflation. In addition to period, we plot the Rossby number ($R_{\rm N}$) of the same two sub-samples of stars against their measured radius inflation in the bottom panel of Figure \ref{fig:period_rossby_fourplot}. Like \citetalias{somers2017}, we found that there is a significant contribution to calculated radius inflation due to binarity. We also found that in considering both rotation period and Rossby number, that the distribution of slowly rotating single stars and binaries are different enough to claim here that we have effectively filtered out binaries from our Hyades sample.

\subsubsection{Distance Errors}\label{sssec:distances}
\par 
The calculation of the stellar radius, which is important as the baseline by which a star is judged to be radially inflated or not is vulnerable to either spurious distance measurements or high errors on the distance itself. High error arises from the typical employment of a cluster distance $\pm$ cluster depth,  where the cluster depth is then taken as the error on the distance of all the stars. For the Pleiades, this was taken to be 134$\pm$3 pc \citep[from][]{soderblom2005}.  

We are able to consider the individual distances to all of our Hyades stars with the {\it Gaia\/} DR2 catalog. The {\it Gaia\/} DR2 catalog measured proper-motions and parallaxes for over $10^{9}$ stars down to a magnitude limit of $M_{G} \sim 20$, with parallax errors approaching 40 $\mu as$. For our final sample of Hyades single stars, we found that the Gaia DR2 parallax errors span a range of [37.8, 511.9] $\mu as$ with an average of 86 $\mu as$. \red{Using the distance derivations in parsecs from \cite{bailor_jones_2018} we have an error on the individual distances of our single stars spanning the range [0.064, 2.435] pc, with an average distance error of 0.30 pc.} We plot the distances and \teff\ for the single stars in our Hyades sample, along with the relevant errors in Figure \ref{fig:hyades_teff_gaia_dist}. 

The plotted single stars are all grouped around the typically quoted cluster distance to the Hyades of 46.75 pc \citep[see][]{reino2018}. The scatter within this group exhibits the cluster depth along our line of sight and is well within the typically cited tidal radius of the Hyades of $\pm$10 pc \citep[][]{reino2018}. From the Figure \ref{fig:hyades_teff_gaia_dist}, as well as the relatively small contribution of distance errors shown in Figure \ref{fig:hyades_frac_errors}, it is apparent that the distance measurements from {\it Gaia\/} DR2 for the Hyades are precise enough to employ for our conversion from angular radius to stellar radius.

\begin{figure*}
%\hspace*{-1.5cm}
% \epsscale{1.2}
\hspace*{-0.5cm}
\includegraphics[scale=1]{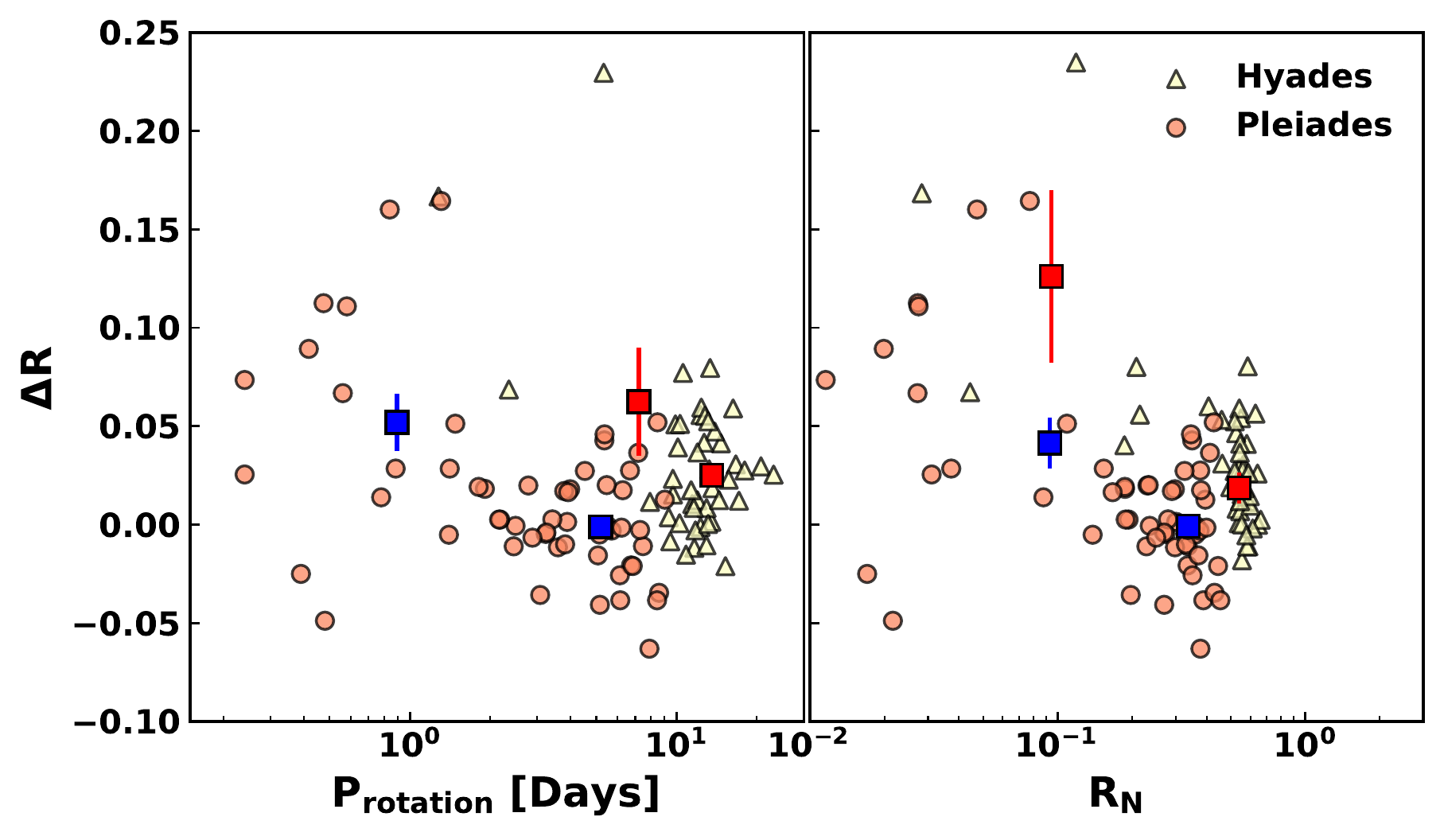}
\begin{center}
\parbox{16cm}{\caption{Right Panel: $\Delta R$ versus rotation period for both the Pleiades (circles) and Hyades (triangles) samples. The blue and red squares correspond to means of the two bins within the Pleiades and Hyades samples respectively. The Hyades is binned into slow rotators ($P > 10d$) and fast rotators ($P < 10$ d). The Pleiades is binned into slow rotators ($P > 2$ d) and fast rotators ($P < 2$ d). Left panel: Plot of Rossby number ($R_{\rm N}$) against $\Delta R$ for both the Hyades (triangles) and Pleiades (circles) single star samples. The two red squares are the mean values of $\Delta R$ for the rapidly convective ($R_{\rm N} < 0.2$) bin and the slowly convective ($R_{\rm N} > 0.2$) bin of Hyades stars. The blue squares are means for the Pleiades stars, binned the same way. }}
\end{center}
\label{fig:rossby_period_delr}
\end{figure*}

\subsection{Radius Inflation and Cluster Evolution: \kgsins{From the Pleiades to the Hyades}}\label{ssec:rossby_rotation}
\par 
We now have two sets of measurements of radius inflation occurring within open clusters of different ages: The Pleiades cluster which has an estimated age of about 125 Myr and the Hyades cluster which has an estimated age of 625 Myr. With this set of data covering a range of about 500 Myr we now have the opportunity to consider radius inflation in the broader sense of cluster evolution. 

\par 
One of the aspects that need to be addressed in comparing the Hyades and Pleiades is the inherent mass difference we find between the samples. The Hyades cluster, being about 625 Myr old has spun down considerably more than the Pleiades. We find the effect of this spin down appear as a lower \teff\ distribution when compared to the younger Pleiades (125 Myr). Our Hyades sample has an average \teff\ of ~4300 K while the Pleiades sample has an average \teff\ of ~5300 K. We test the similarity of these two distributions using the two sample KS test, finding a non-random difference between the two samples. We argue that this does not affect our results as presented, as we are comparing the overall presence of radius inflation as a function of cluster evolution here and not the relative degree of radius inflation.

\par 
%The most conspicuous aspect about radius inflation we find in our results is the suggestion that the process of radius inflation occurs in spite of the spin-down that occurs as stars age. As we showed with Figure \ref{fig:period_rossby_twoplot}, there isn't a significant trend with rotation period and radius inflation within the Hyades. Overall, the Hyades cluster stars spin a lot more slowly than the Pleiades stars. Considering that the Rossby number - radius inflation trend is significant within our results further suggests that the primary mechanism of radius inflation within older clusters is the magnetic field activity that occurs within the convective layers of magnetically active stars.

\par 
%We plot Rossby number, and period against \teff\ and colored by our radius inflation values for the single stars within the Pleiades, and the Hyades within Figure \ref{fig:period_rossby_twoplot2}. From the left panel of Figure \ref{fig:period_rossby_twoplot2}, we show that the more radially inflated stars within both the older Hyades cluster (Triangles), and the younger Pleiades cluster (circles), are generally found below Rossby number = 0.2. 

In the right panel of Fig. \ref{fig:period_rossby_twoplot2}, we plot the rotation rate of single stars in both the Pleiades (circles) and Hyades (triangles) against their \teff. Each datum is colored according to its inferred radius inflation. As shown before, the stars with convincing evidence of inflation (darkest points) are rotating faster than average for their mass range. However, the inflated Hyades stars rotate slower than the inflated Pleiades stars. In the left panel, we show the same plot with the Rossby number substituted for rotation rate. As low-mass stars have a longer convective overturn timescale, the Rossby numbers of the inflated Hyades are similar to those of the faster-rotating Pleiads. In this panel, the inflated members of both cluster now occupy a similar range, namely $\lesssim 0.2$. That the onset of inflation seems more closely tied to Rossby number than raw rotation rate strongly suggests a link with magnetic properties.

An alternative way to view \red{these} data is shown in Figure \ref{fig:rossby_period_delr}. Here we plot the rotation rate and Rossby number of each star directly against the inferred radius inflation. While there seems to exist a connection between rotation and inflation in both clusters, in the sense that below a threshold rotation rate stars can show significant inflation, the transition rotation rate does not align for the two clusters. However, in Rossby space we see again an apparent transition around R$_N \sim 0.2$, above which stars cluster around a low $\Delta R$ value and above which stars may show large radius inflation values. The most likely explanation for this alignment is that the same magnetic mechanism is operating in both clusters, but sets in at different mass thresholds given the progressive spindown of stars with age.

Many previous studies have considered correlations between Rossby number and the magnetic properties of stars \citep{noyes1984,pizzolato2003,wright2011,douglas2014}. A notable result is that around a Rossby number of 0.1--0.2, there is a break in correlation between rotation rate and magnetic proxies (e.g., H$\alpha$, X-rays, Ca~II H\&K). Towards larger Rossby numbers, the magnetic proxies decrease in strength with decreasing rotation rate, but at lower Rossby numbers the correlation ceases and the strength of magnetic proxies are approximately constant regardless of the rotation rate. A similar connection between Rossby number and starspot properties has also been noted \citep{odell1995} SS17 noted that radius inflation set in around this Rossby value in the Pleiades, suggesting a connection. Figure~\ref{fig:rossby_period_delr} shows that this Rossby threshold appears to hold in the much older Hyades as well, even though the low-Rossby stars are in a different mass range. This strongly suggests that the still-unclear physics of magnetic saturation are related to the onset of radius inflation at high rotation rate. 

%There has been previous works that explored the Rossby number as a proxy for magnetic activity within the convective layers of stars.  [CITATION, Newton?], and [CITATION, Pizzolato?] looked at bolometric luminosity and Rossby number, finding a threshold occurring around 0.1. This threshold is to distinguish between stars that are saturated magnetically (R $<$ 0.1) and unsaturated (R $>$ 0.1) [ADD MORE ABOUT THE ACTUAL PHYSICS TAKING PLACE THAT RESULT IN THIS]. As we find a similar threshold in Rossby number occuring when considering radius inflation in the Hyades, as well as the Pleiades, this suggests that the magnetic activity generated within a star's convective layers continues despite the spin-down that stars undergo as a cluster ages.

\par 
%The spin-down in rotational velocity of stars as they age and its connection to magnetic activity has been explored before. [Skumanich (1972)], considering Ca II emission data for the Hyades, Pleiades, and Ursa Major, found results suggesting that both rotational velocity, and Ca II emission deteriorate in proportion to the inverse square root of the age of the cluster. Ca II emission here served as an analog for the average magnetic field strength of a star.

%\textbf{[]Having trouble composing discussion level arguments about the aging clusters and their spin down rates and how it relates to radii inflation without sounding like I'm going in circles.]}

\section{conclusions}\label{sec:conclusions}
\par 

The causes of radius inflation---the tendency of some low mass stars to be physically larger by $\sim$10\% than both theoretical predictions and other stars of equal mass, age, and composition---remains an active area of research. In a recent paper, \citetalias{somers2017} searched for the presence of radius inflation among the low mass members of the Pleiades, finding a significant correlation between rapid rotation and radius inflation. This strongly implicated dynamo-generated magnetic fields, which are strong in rapidly-rotating young stars, as the driver of inflation. In this work, we have expanded the reach of this investigation to the older Hyades cluster in order to test how the inflation pattern within clusters evolves as a function of age. 

We used archival photometry, distances, and rotation rates for a sample of 68 single-star Hyads between $\sim 0.5-1$~\msun, and empirically determined their \teffs\ and radii. By assuming a correlation between inflated radius and reduced surface temperature (see $\S$\ref{ssec:radii_inflation}) we calculated the fractional deviation of each star's radius from theoretical expectations. From this exercise, we found several Hyades members exhibiting radius inflation. 

These values reveal a strong correlation between the degree of radius inflation and Rossby number (R$_{\rm N}$), a robust indicator of the strength of dynamo-generated magnetic fields. Thus we concluded that Hyades members with R$_{\rm N}$ below 0.1 (implying very rapid rotation) tend to be physically larger than their siblings with higher R$_{\rm N}$s (slower rotation). 

Finally, we compared these Hyades results to the findings of \citetalias{somers2017} for the Pleiades. Those authors only reported a correlation with rotation rate, so we re-analyzed their results with the methods laid out in this paper. We find that the \citetalias{somers2017} Pleiades sample also shows a strong correlation between radius inflation and R$_{\rm N}$. Moreover, and remarkably, the Pleiades exhibits the same transition point at R$_{\rm N} \sim 0.1$ as the Hyades, with inflated members below this value and non-inflated members above. This strongly suggests that this R$_{\rm N}$ threshold for radius inflation is universal, at least over the range of cluster ages spanned by the Pleiades ($\sim$100~Myr) to the Hyades ($\sim$700~Myr). Future work should focus on characterizing inflation over a larger range of open cluster ages, in order to assess the full age domain in which this Rossby-inflation correlation persists.

Notably, the mass range of stars with low R$_{\rm N}$ is lower in the Hyades than in the Pleiades due to the progressive spin-down of stars below $\sim 1.3$\msun\ as they age. This implicates rotation, and not the mass range, as the primary corollary of radius inflation. More fundamentally, because the canonical rotation-activity relation of low-mass stars is understood to result from the connection between magnetic activity and surface convection, our results imply that magnetic activity within the convective layers of low-mass stars is what preferentially drives radius inflation.

\acknowledgements
We thank the referee for their comments and suggestions which greatly improved the quality and scope of this paper. KOJ and GS acknowledge the support of the Vanderbilt Office of the Provost through the Vanderbilt Initiative in Data-intensive Astrophysics (VIDA). KOJ thanks the LSSTC Data Science Fellowship Program, which is funded by LSSTC, NSF Cybertraining Grant \#1829740, the Brinson Foundation, and the Moore Foundation; their participation in the program has benefited this work. KGS acknowledges NASA grant 17-XRP17 2-0024 for partial support. This research made use of the cross-match service provided by CDS, Strasbourg.

\clearpage
\begin{deluxetable*}{ccccccccc}
\tablecaption{Observed Stellar Properties \label{tab:basic_stellar_data}}
\tablewidth{0pt}
\tablehead{
\colhead{2MASS}
&\colhead{R.A.}
&\colhead{Decl.}
&\colhead{Distance}
&\colhead{V$_{\rm mag}$}
&\colhead{$B -V$}
&\colhead{$V - K_{S}$}
&\colhead{$P_{\rm rot}$}
&\colhead{Binary?}\\
\colhead{}
&\colhead{}
&\colhead{}
&\colhead{(PC)}
&\colhead{}
&\colhead{}
&\colhead{}
&\colhead{(Days)}
&\colhead{}
} 
\startdata
03373495+2120355&54.3956&21.3432&45.81$^{+0.13}_{-0.13}$&9.362$\pm$0.008&0.92$\pm$0.004&2.240$\pm$0.021&10.57&No     \\
03434706+2051363&55.946&20.86&45.21$^{+0.13}_{-0.13}$&14.54$\pm$0.008&0.9$\pm$0.004&4.927$\pm$0.023&12.3&No     \\
03510309+2354134&57.7628&23.9037&40.83$^{+0.08}_{-0.08}$&10.118$\pm$0.008&1.2819$\pm$0.004&2.723$\pm$0.021&12.57&Yes     \\
03524101+2548159&58.1708&25.8044&44.41$^{+0.12}_{-0.12}$&11.21$\pm$0.008&  &3.317$\pm$0.027&14.66&No     \\
03550142+1229081&58.7559&12.4855&45.98$^{+0.09}_{-0.09}$&10.094$\pm$0.008&1.07$\pm$0.004&2.519$\pm$0.022&11.66&No     \\
03583581+1306182&59.6492&13.105&61.24$^{+0.19}_{-0.19}$&8.951$\pm$0.008&  &1.576$\pm$0.022&22.26&Yes?     \\
03590972+2628340&59.7905&26.4761&35.82$^{+0.09}_{-0.09}$&11.47$\pm$0.008&  &3.585$\pm$0.021&15.25&No     \\
03591417+2202380&59.809&22.0439&40.55$^{+0.08}_{-0.08}$&13.066$\pm$0.008&  &4.362$\pm$0.018&20.73&No     \\
04033902+1927180&60.9125&19.455&47.97$^{+0.11}_{-0.11}$&10.095$\pm$0.008&1.08$\pm$0.004&2.495$\pm$0.021&11.45&No     \\
04052565+1926316&61.3568&19.4421&47.23$^{+0.12}_{-0.12}$&11.4$\pm$0.008&1.35$\pm$0.004&3.292$\pm$0.027&13.51&No     \\
04063463+1332566&61.6443&13.549&49.84$^{+0.38}_{-0.38}$&13.409$\pm$0.008&1.47$\pm$0.004&4.318$\pm$0.019&16.68&No     \\
04070122+1520062&61.7551&15.335&45.03$^{+0.24}_{-0.24}$&10.473$\pm$0.008&1.18$\pm$0.004&2.808$\pm$0.027&11.98&No     \\
04074319+1631076&61.9299&16.5187&46.5$^{+0.11}_{-0.11}$&9.924$\pm$0.008&1.02$\pm$0.004&2.412$\pm$0.027&12.3&No     \\
04081110+1652229&62.0462&16.873&40.09$^{+0.07}_{-0.07}$&11.51$\pm$0.008&1.44$\pm$0.004&3.577$\pm$0.025&13.63&No     \\
04082667+1211304&62.1111&12.1918&46.38$^{+0.15}_{-0.15}$&11.269$\pm$0.008&1.33$\pm$0.004&3.223$\pm$0.028&12.96&No     \\
04083620+2346071&62.1508&23.7686&47.45$^{+0.13}_{-0.12}$&9.41$\pm$0.008&0.9$\pm$0.004&2.087$\pm$0.016&9.35&No     \\
04084015+2333257&62.1673&23.5571&46.02$^{+0.17}_{-0.17}$&12.861$\pm$0.008&  &4.332$\pm$0.025&21.02&Yes     \\
04115620+2338108&62.9841&23.6363&40.3$^{+1.19}_{-1.13}$&9.392$\pm$0.008&  &2.982$\pm$0.025&2.309&Yes     \\
04151038+1423544&63.7932&14.3984&47.95$^{+0.09}_{-0.09}$&11.585$\pm$0.008&1.38$\pm$0.004&3.457$\pm$0.018&13.91&No     \\
04153367+1542226&63.8902&15.7062&45.52$^{+0.14}_{-0.14}$&10.931$\pm$0.008&  &3.077$\pm$0.024&47&Yes     \\
04163346+2154269&64.1394&21.9074&51.02$^{+0.11}_{-0.11}$&9.125$\pm$0.008&0.81$\pm$0.004&1.849$\pm$0.024&10.26&No     \\
04172512+1901478&64.3547&19.0299&47.79$^{+0.18}_{-0.18}$&10.8$\pm$0.008&1.22$\pm$0.004&2.891$\pm$0.024&12.95&No     \\
04172811+1454038&64.3671&14.901&49.45$^{+0.13}_{-0.13}$&14.47$\pm$0.008&1.55$\pm$0.004&4.849$\pm$0.020&2.35&No     \\
04174767+1339422&64.4486&13.6617&47.39$^{+0.12}_{-0.12}$&12.54$\pm$0.008&  &4.012$\pm$0.020&8.685&Yes     \\
04175061+1828307&64.4609&18.4751&46.65$^{+0.31}_{-0.31}$&13.954$\pm$0.008&  &4.962$\pm$0.019&22.94&Yes     \\
04175555+1432464&64.4814&14.5462&52.38$^{+0.2}_{-0.2}$&14.89$\pm$0.008&  &4.873$\pm$0.018&10.11&No     \\
04181077+2317048&64.5449&23.2846&53.92$^{+0.13}_{-0.13}$&9.471$\pm$0.008&  &2.534$\pm$0.017&1.862&Yes     \\
04223953+1816097&65.6647&18.2693&41.66$^{+0.09}_{-0.09}$&13.04$\pm$0.008&  &4.313$\pm$0.019&23.12&No     \\
04232283+1939312&65.8451&19.6586&45.71$^{+0.34}_{-0.34}$&9.381$\pm$0.008&0.91$\pm$0.004&2.230$\pm$0.021&9.9&No     \\
04232526+1545474&65.8552&15.7631&41.47$^{+0.15}_{-0.15}$&10.472$\pm$0.008&1.24$\pm$0.004&2.977$\pm$0.019&12.38&No     \\
04235070+0912193&65.9612&9.20538&44.83$^{+0.11}_{-0.11}$&12.896$\pm$0.008&1.51$\pm$0.004&4.67$\pm$0.019&5.33&No     \\
04235911+1643178&65.9963&16.7216&46.0$^{+0.12}_{-0.12}$&12.514$\pm$0.008&1.49$\pm$0.004&3.952$\pm$0.019&17.14&No     \\
04241691+1800107&66.0704&18.0029&46.66$^{+0.09}_{-0.09}$&9.966$\pm$0.008&1.06$\pm$0.004&2.441$\pm$0.019&11.6&No     \\
04250024+1659057&66.251&16.9849&54.61$^{+0.19}_{-0.19}$&10.248$\pm$0.008&1.03$\pm$0.004&2.417$\pm$0.017&11.77&No     \\
04251456+1858250&66.3106&18.9736&52.87$^{+0.2}_{-0.2}$&12.728$\pm$0.008&1.48$\pm$0.004&4.022$\pm$0.019&10.84&Yes     \\
04252501+1754552&66.3542&17.9153&46.85$^{+0.09}_{-0.09}$&11.128$\pm$0.008&  &3.151$\pm$0.019&70&Yes     \\
04254754+1801022&66.448&18.0172&42.11$^{+1.05}_{-1.0}$&8.9889$\pm$0.008&0.94$\pm$0.004&2.221$\pm$0.022&8.46&Yes     \\
04264825+1052160&66.701&10.8711&46.97$^{+1.01}_{-0.97}$&9.432$\pm$0.008&1.04$\pm$0.004&2.549$\pm$0.021&10.4&Yes     \\
04272532+1415384&66.8555&14.2606&52.59$^{+0.51}_{-0.5}$&10.313$\pm$0.008&1.08$\pm$0.004&2.600$\pm$0.024&12.77&No     \\
04274701+1425041&66.9459&14.4178&49.32$^{+0.1}_{-0.1}$&9.475$\pm$0.008&0.92$\pm$0.004&2.127$\pm$0.021&9.7&No     \\
04275895+1830009&66.9956&18.5002&51.06$^{+0.27}_{-0.27}$&10.129$\pm$0.008&  &2.596$\pm$0.019&11.13&Yes     \\
04282878+1741453&67.1199&17.6959&46.83$^{+0.58}_{-0.57}$&12.122$\pm$0.008&1.49$\pm$0.004&4.411$\pm$0.027&2.42&Yes     \\
04303385+1444532&67.641&14.7481&49.98$^{+1.17}_{-1.12}$&14.72$\pm$0.008&1.56$\pm$0.004&5.008$\pm$0.019&18.41&Yes     \\
04303486+1544023&67.6452&15.7339&57.07$^{+0.9}_{-0.87}$&8.84$\pm$0.008&0.84$\pm$0.004&2.022$\pm$0.019&8.73&Yes     \\
04315244+1529585&67.9685&15.4995&45.73$^{+0.08}_{-0.08}$&11.0$\pm$0.008&1.31$\pm$0.004&3.105$\pm$0.025&13.13&No     \\
04322565+1306476&68.1068&13.1132&46.72$^{+0.1}_{-0.1}$&11.0$\pm$0.008&1.19$\pm$0.004&3.346$\pm$0.021&1.48&Yes     \\
04332699+1302438&68.3624&13.0455&43.09$^{+0.1}_{-0.1}$&13.328$\pm$0.008&1.57$\pm$0.004&4.505$\pm$0.020&16.29&No     \\
04333716+2109030&68.4048&21.1508&43.74$^{+0.09}_{-0.09}$&10.726$\pm$0.008&1.23$\pm$0.004&3.040$\pm$0.027&12.69&No     \\
04341113+1133285&68.5464&11.5579&47.78$^{+0.5}_{-0.49}$&11.25$\pm$0.008&1.39$\pm$0.004&3.191$\pm$0.030&11.03&Yes     \\
04354850+1317169&68.9521&13.288&49.17$^{+0.14}_{-0.14}$&14.91$\pm$0.008&1.63$\pm$0.004&5.042$\pm$0.019&13.36&No     \\
04360525+1541026&69.0218&15.684&50.15$^{+0.12}_{-0.12}$&9.345$\pm$0.008&0.87$\pm$0.004&1.972$\pm$0.016&9.47&No     \\
04395095+1243426&69.9623&12.7285&43.63$^{+0.09}_{-0.09}$&9.992$\pm$0.008&1.07$\pm$0.004&2.512$\pm$0.024&10.85&No     \\
04412780+1404340&70.3658&14.0761&49.2$^{+1.28}_{-1.22}$&13.395$\pm$0.008&  &4.683$\pm$0.022&1.28&No     \\
04412876+1200337&70.3698&12.0093&47.16$^{+0.1}_{-0.09}$&12.898$\pm$0.008&1.5$\pm$0.004&4.144$\pm$0.021&18&No     \\
04431568+1704088&70.8153&17.0691&45.88$^{+0.41}_{-0.4}$&9.92$\pm$0.008&1.0$\pm$0.004&2.524$\pm$0.017&10.31&No     \\
04461879+0338108&71.5783&3.63636&45.1$^{+0.08}_{-0.08}$&10.922$\pm$0.008&1.27$\pm$0.004&3.096$\pm$0.021&13.25&No     \\
04463036+1528194&71.6265&15.472&49.05$^{+0.21}_{-0.2}$&8.28$\pm$0.008&0.66$\pm$0.004&1.541$\pm$0.033&7.95&No     \\
04471851+0627113&71.8271&6.45315&40.54$^{+0.07}_{-0.06}$&11.348$\pm$0.008&1.42$\pm$0.004&3.473$\pm$0.027&14.44&No     \\
04480086+1703216&72.0036&17.056&44.13$^{+0.9}_{-0.87}$&11.085$\pm$0.008&1.41$\pm$0.004&3.711$\pm$0.024&10.77&Yes     \\
04483062+1623187&72.1276&16.3885&48.12$^{+0.13}_{-0.13}$&12.427$\pm$0.008&1.47$\pm$0.004&3.884$\pm$0.018&15.69&No     \\
04484211+2106035&72.1754&21.1009&45.37$^{+0.23}_{-0.23}$&9.057$\pm$0.008&0.85$\pm$0.004&1.985$\pm$0.019&9.69&No     \\
04491296+2448103&72.304&24.8028&49.59$^{+0.14}_{-0.14}$&9.492$\pm$0.008&1.04$\pm$0.004&2.536$\pm$0.038&6.9&Yes     \\
04500069+1624436&72.5028&16.4121&49.34$^{+0.13}_{-0.13}$&10.69$\pm$0.008&1.16$\pm$0.004&2.851$\pm$0.017&11.98&No     \\
04510241+1458167&72.76&14.9713&40.77$^{+0.75}_{-0.72}$&11.675$\pm$0.008&  &3.708$\pm$0.021&13.14&No     \\
04522352+1859489&73.098&18.9969&50.75$^{+0.12}_{-0.12}$&10.28$\pm$0.008&1.07$\pm$0.004&2.537$\pm$0.021&11.34&No     \\
04522385+1043099&73.0994&10.7194&52.04$^{+0.14}_{-0.14}$&12.816$\pm$0.008&  &4.044$\pm$0.023&9.88&Yes?     \\
05054038+0627545&76.4182&6.46515&65.44$^{+1.43}_{-1.37}$&9.88$\pm$0.008&0.95$\pm$0.004&2.362$\pm$0.022&10.41&Yes     \\
05110971+1548574&77.7904&15.8159&57.09$^{+2.44}_{-2.25}$&12.08$\pm$0.008&  &3.977$\pm$0.018&14.94&Yes?     \\
\enddata
\end{deluxetable*}

\begin{deluxetable*}{cccccccc}
% \tabletypesize{\scriptsize}
\tablecaption{Derived Stellar Properties ($V-K_s$) \label{tab:derived_stellar_data}}
% \tablewidth{0pt}
\tablehead{
\colhead{2MASS}
&\colhead{\teff}
&\colhead{$\sigma_{teff}$}
&\colhead{$\mathcal{F}_{bol}$}
&\colhead{$\chi_{\nu}^2$}
&\colhead{Angular Diameter}
&\colhead{Radius}
&\colhead{$\Delta$Radius}\\
\colhead{}
&\colhead{}
&\colhead{(Systematic)}
&\colhead{$\times 10^{-10}$}
&\colhead{}
&\colhead{$\times 10^{-2}$}
&\colhead{}
&\colhead{}\\
\colhead{}
&\colhead{(K)}
&\colhead{(K)}
&\colhead{(erg cm$^{-2}$ s$^{-1}$)}
&\colhead{}
&\colhead{(mas)}
&\colhead{($R_{\odot})$}
&\colhead{(\%)}
} 
\startdata
3373495+2120355&4967$\pm$14.29&60&62.8$^{+2.0}_{-2.52}$&1.22&8.8$^{+0.15}_{-0.18}$&0.87$^{+0.01}_{-0.02}$&7.7108$^{+3.663}_{-3.663}$   \\
3434706+2051363&3419$\pm$6.36&60&2.67$^{+0.245}_{-0.215}$&7.21&3.83$^{+0.18}_{-0.15}$&0.37$^{+0.02}_{-0.02}$&5.5743$^{+6.351}_{-6.351}$   \\
3510309+2354134&4565$\pm$11.71&60&36.5$^{+1.3599}_{-0.975}$&2.8&7.94$^{+0.15}_{-0.11}$&0.7$^{+0.01}_{-0.01}$&-0.044$^{+3.655}_{-3.655}$   \\
3524101+2548159&4169$\pm$12.35&60&19.6$^{+0.571}_{-0.547}$&0.95&6.98$^{+0.11}_{-0.11}$&0.67$^{+0.01}_{-0.01}$&4.1394$^{+3.929}_{-3.929}$   \\
3550142+1229081&4725$\pm$13.34&60&34.5$^{+1.97}_{-1.820}$&5.98&7.2$^{+0.21}_{-0.19}$&0.71$^{+0.02}_{-0.02}$&-1.172$^{+4.171}_{-4.171}$   \\
3583581+1306182&5686$\pm$20.52&60&71.5$^{+3.6400}_{-3.39}$&6.11&7.16$^{+0.19}_{-0.18}$&0.94$^{+0.03}_{-0.02}$&-0.048$^{+3.818}_{-3.818}$   \\
3590972+2628340&4012$\pm$8.59&60&18.2$^{+0.364}_{-0.527}$&1.18&7.26$^{+0.08}_{-0.11}$&0.56$^{+0.01}_{-0.01}$&-2.101$^{+3.496}_{-3.496}$   \\
3591417+2202380&3637$\pm$5.89&60&7.090$^{+0.4479}_{-0.458}$&8.71&5.51$^{+0.18}_{-0.18}$&0.48$^{+0.02}_{-0.02}$&2.9676$^{+4.954}_{-4.954}$   \\
4033902+1927180&4745$\pm$12.84&60&34.9$^{+1.66}_{-1.55}$&2.5&7.19$^{+0.18}_{-0.16}$&0.74$^{+0.02}_{-0.02}$&1.0510$^{+3.930}_{-3.930}$   \\
4052565+1926316&4183$\pm$12.45&60&15.5$^{+0.552}_{-0.524}$&2.04&6.16$^{+0.12}_{-0.11}$&0.63$^{+0.01}_{-0.01}$&0.1523$^{+3.908}_{-3.908}$   \\
04063463+1332566&3656$\pm$6.25&60&5.020$^{+0.29}_{-0.303}$&6.28&4.59$^{+0.13}_{-0.14}$&0.49$^{+0.01}_{-0.02}$&3.0635$^{+4.840}_{-4.840}$   \\
04070122+1520062&4506$\pm$14.91&60&28.80$^{+0.8029}_{-0.771}$&0.49&7.24$^{+0.11}_{-0.11}$&0.7$^{+0.01}_{-0.01}$&1.2427$^{+3.690}_{-3.690}$   \\
04074319+1631076&4819$\pm$17.5&60&39.5$^{+1.44}_{-1.69}$&5.32&7.41$^{+0.15}_{-0.17}$&0.74$^{+0.01}_{-0.02}$&-0.111$^{+3.700}_{-3.700}$   \\
04081110+1652229&4018$\pm$10.36&60&17.1$^{+0.8260}_{-0.917}$&5.33&7.01$^{+0.17}_{-0.19}$&0.61$^{+0.01}_{-0.02}$&1.8593$^{+4.336}_{-4.336}$   \\
04082667+1211304&4227$\pm$13.31&60&17.7$^{+0.33}_{-0.162}$&0.29&6.45$^{+0.07}_{-0.05}$&0.64$^{+0.01}_{-0.01}$&0.8483$^{+3.636}_{-3.636}$   \\
04083620+2346071&5109$\pm$11.96&60&55.89$^{+1.56}_{-1.49}$&1.55&7.84$^{+0.12}_{-0.11}$&0.8$^{+0.01}_{-0.01}$&0.3837$^{+3.166}_{-3.166}$   \\
04084015+2333257&3652$\pm$8.17&60&8.68$^{+0.534}_{-0.489}$&4.8&6.05$^{+0.19}_{-0.17}$&0.6$^{+0.02}_{-0.02}$&11.581$^{+5.412}_{-5.412}$   \\
04115620+2338108&4381$\pm$12.79&60&88.1$^{+2.73}_{-3.87}$&2.85&13.39$^{+0.22}_{-0.3}$&1.16$^{+0.04}_{-0.04}$&41.846$^{+6.679}_{-6.679}$   \\
04151038+1423544&4081$\pm$7.75&60&14.80$^{+0.669}_{-0.381}$&2.83&6.32$^{+0.14}_{-0.08}$&0.65$^{+0.02}_{-0.01}$&4.7336$^{+4.210}_{-4.210}$   \\
04153367+1542226&4317$\pm$11.8&60&21.29$^{+1.16}_{-1.08}$&7.22&6.78$^{+0.19}_{-0.18}$&0.66$^{+0.02}_{-0.02}$&1.0256$^{+4.354}_{-4.354}$   \\
04163346+2154269&5365$\pm$19.71&60&66.8$^{+1.29}_{-2.46}$&4.05&7.78$^{+0.09}_{-0.15}$&0.85$^{+0.01}_{-0.02}$&0.0777$^{+3.134}_{-3.134}$   \\
04172512+1901478&4445$\pm$12.66&60&21.8$^{+0.9770}_{-0.915}$&6.27&6.47$^{+0.15}_{-0.14}$&0.67$^{+0.02}_{-0.01}$&-1.043$^{+3.939}_{-3.939}$   \\
04172811+1454038&3446$\pm$5.68&60&2.900$^{+0.154}_{-0.123}$&4.14&3.93$^{+0.11}_{-0.08}$&0.42$^{+0.01}_{-0.01}$&6.8703$^{+4.971}_{-4.971}$   \\
04174767+1339422&3794$\pm$7.16&60&8.690$^{+0.581}_{-0.528}$&8.58&5.61$^{+0.19}_{-0.17}$&0.57$^{+0.02}_{-0.02}$&4.8220$^{+5.110}_{-5.110}$   \\
04175061+1828307&3405$\pm$5.28&60&4.979$^{+0.287}_{-0.264}$&5.82&5.27$^{+0.15}_{-0.14}$&0.53$^{+0.02}_{-0.01}$&16.098$^{+5.621}_{-5.621}$   \\
04175555+1432464&3437$\pm$5.16&60&1.870$^{+0.1750}_{-0.037}$&7.03&3.17$^{+0.15}_{-0.03}$&0.36$^{+0.02}_{-0.0}$&3.9298$^{+6.272}_{-6.272}$   \\
04181077+2317048&4710$\pm$10.37&60&66.1$^{+2.06}_{-1.97}$&0.81&10.04$^{+0.16}_{-0.16}$&1.17$^{+0.02}_{-0.02}$&33.526$^{+4.510}_{-4.510}$   \\
04223953+1816097&3658$\pm$6.26&60&7.02$^{+0.3}_{-0.282}$&3.41&5.42$^{+0.12}_{-0.11}$&0.49$^{+0.01}_{-0.01}$&2.5476$^{+4.319}_{-4.319}$   \\
04232283+1939312&4977$\pm$14.36&60&59.0$^{+2.0100}_{-3.13}$&8.35&8.49$^{+0.15}_{-0.23}$&0.84$^{+0.02}_{-0.02}$&5.1025$^{+3.700}_{-3.700}$   \\
04232526+1545474&4381$\pm$9.63&60&32.8$^{+0.6310}_{-0.912}$&0.6&8.17$^{+0.09}_{-0.12}$&0.73$^{+0.01}_{-0.01}$&5.9528$^{+3.538}_{-3.538}$   \\
04235070+0912193&3513$\pm$5.68&60&9.959$^{+0.7859}_{-0.591}$&8.15&7.0$^{+0.28}_{-0.21}$&0.68$^{+0.03}_{-0.02}$&22.958$^{+6.690}_{-6.690}$   \\
04235911+1643178&3822$\pm$6.95&60&8.42$^{+0.66}_{-0.591}$&5.06&5.44$^{+0.21}_{-0.19}$&0.54$^{+0.02}_{-0.02}$&1.2178$^{+5.327}_{-5.327}$   \\
04241691+1800107&4788$\pm$11.88&60&38.8$^{+0.6990}_{-0.679}$&0.65&7.44$^{+0.08}_{-0.07}$&0.75$^{+0.01}_{-0.01}$&0.8544$^{+3.168}_{-3.168}$   \\
04250024+1659057&4806$\pm$10.87&60&28.00$^{+1.01}_{-0.954}$&1.91&6.27$^{+0.12}_{-0.11}$&0.74$^{+0.01}_{-0.01}$&-0.287$^{+3.463}_{-3.463}$   \\
04251456+1858250&3789$\pm$6.81&60&7.25$^{+0.54}_{-0.485}$&4.89&5.14$^{+0.19}_{-0.17}$&0.58$^{+0.02}_{-0.02}$&6.0036$^{+5.451}_{-5.451}$   \\
04252501+1754552&4267$\pm$9.04&60&17.9$^{+0.9440}_{-1.01}$&8.25&6.36$^{+0.17}_{-0.18}$&0.64$^{+0.02}_{-0.02}$&-0.111$^{+4.173}_{-4.173}$   \\
04254754+1801022&4985$\pm$15.13&60&86.2$^{+3.0900}_{-4.8}$&3.69&10.23$^{+0.19}_{-0.29}$&0.93$^{+0.03}_{-0.03}$&11.693$^{+4.799}_{-4.799}$   \\
04264825+1052160&4700$\pm$12.55&60&67.1$^{+1.63}_{-3.82}$&6.11&10.16$^{+0.13}_{-0.29}$&1.03$^{+0.03}_{-0.04}$&24.501$^{+4.923}_{-4.923}$   \\
04272532+1415384&4662$\pm$14.19&60&29.8$^{+0.532}_{-0.518}$&0.91&6.88$^{+0.07}_{-0.07}$&0.78$^{+0.01}_{-0.01}$&5.5282$^{+3.632}_{-3.632}$   \\
04274701+1425041&5075$\pm$15.03&60&51.4$^{+0.9609}_{-1.39}$&2.81&7.62$^{+0.08}_{-0.11}$&0.81$^{+0.01}_{-0.01}$&1.5360$^{+3.155}_{-3.155}$   \\
04275895+1830009&4662$\pm$11.16&60&35.8$^{+1.06}_{-1.02}$&2.62&7.54$^{+0.12}_{-0.11}$&0.83$^{+0.01}_{-0.01}$&9.7850$^{+3.769}_{-3.769}$   \\
04282878+1741453&3620$\pm$8.69&60&16.9$^{+1.0}_{-0.920}$&5.68&8.59$^{+0.26}_{-0.24}$&0.87$^{+0.03}_{-0.03}$&36.433$^{+6.779}_{-6.779}$   \\
04303385+1444532&3389$\pm$5.22&60&2.59$^{+0.134}_{-0.124}$&4.94&3.84$^{+0.1}_{-0.09}$&0.41$^{+0.01}_{-0.01}$&9.2007$^{+5.673}_{-5.673}$   \\
04303486+1544023&5176$\pm$14.26&60&87.4$^{+3.15}_{-2.99}$&8.41&9.56$^{+0.18}_{-0.17}$&1.17$^{+0.03}_{-0.03}$&22.461$^{+4.575}_{-4.575}$   \\
04315244+1529585&4300$\pm$12.22&60&19.9$^{+0.771}_{-0.904}$&8.11&6.61$^{+0.13}_{-0.15}$&0.65$^{+0.01}_{-0.02}$&0.0410$^{+3.882}_{-3.882}$   \\
04322565+1306476&4148$\pm$9.31&60&22.9$^{+1.5}_{-1.150}$&5.87&7.62$^{+0.25}_{-0.19}$&0.77$^{+0.03}_{-0.02}$&13.762$^{+5.328}_{-5.328}$   \\
04332699+1302438&3578$\pm$6.21&60&6.11$^{+0.3500}_{-0.322}$&5.88&5.29$^{+0.15}_{-0.14}$&0.49$^{+0.01}_{-0.01}$&5.9090$^{+4.961}_{-4.961}$   \\
04333716+2109030&4344$\pm$13.65&60&26.5$^{+0.758}_{-0.962}$&2.3&7.47$^{+0.12}_{-0.14}$&0.7$^{+0.01}_{-0.01}$&4.1654$^{+3.839}_{-3.839}$   \\
04341113+1133285&4248$\pm$14.58&60&17.80$^{+0.7159}_{-0.838}$&1.99&6.4$^{+0.14}_{-0.16}$&0.66$^{+0.02}_{-0.02}$&1.7683$^{+4.250}_{-4.250}$   \\
04354850+1317169&3377$\pm$nan&60&2.199$^{+0.147}_{-0.133}$&6.86&3.56$^{+nan}_{-nan}$&0.38$^{+nan}_{-nan}$&7.9629$^{+nan}_{-nan}$   \\
04360525+1541026&5225$\pm$12.6&60&55.79$^{+1.05}_{-2.02}$&4.13&7.49$^{+0.08}_{-0.14}$&0.81$^{+0.01}_{-0.02}$&-0.844$^{+2.918}_{-2.918}$   \\
04395095+1243426&4733$\pm$14.7&60&38.3$^{+0.684}_{-0.993}$&3.26&7.56$^{+0.08}_{-0.11}$&0.71$^{+0.01}_{-0.01}$&-1.518$^{+3.237}_{-3.237}$   \\
04412780+1404340&3509$\pm$6.49&60&6.42$^{+0.4870}_{-0.437}$&4.78&5.63$^{+0.21}_{-0.19}$&0.6$^{+0.03}_{-0.03}$&16.672$^{+6.953}_{-6.953}$   \\
04412876+1200337&3733$\pm$7.21&60&6.88$^{+0.468}_{-0.425}$&9.4&5.15$^{+0.18}_{-0.16}$&0.52$^{+0.02}_{-0.02}$&2.7593$^{+5.093}_{-5.093}$   \\
04431568+1704088&4718$\pm$10.41&60&41.7$^{+2.59}_{-1.609}$&2.13&7.95$^{+0.25}_{-0.16}$&0.78$^{+0.03}_{-0.02}$&5.1217$^{+4.621}_{-4.621}$   \\
04461879+0338108&4303$\pm$10.17&60&22.5$^{+0.634}_{-0.408}$&0.55&7.01$^{+0.1}_{-0.07}$&0.68$^{+0.01}_{-0.01}$&2.8151$^{+3.657}_{-3.657}$   \\
04463036+1528194&5745$\pm$34.28&60&127.0$^{+5.5100}_{-1.32}$&1.61&9.35$^{+0.23}_{-0.12}$&0.99$^{+0.02}_{-0.01}$&1.1597$^{+4.002}_{-4.002}$   \\
04471851+0627113&4077$\pm$11.7&60&18.2$^{+1.12}_{-1.02}$&5.98&7.03$^{+0.22}_{-0.2}$&0.61$^{+0.02}_{-0.02}$&1.2485$^{+4.733}_{-4.733}$   \\
04480086+1703216&3946$\pm$9.48&60&26.0$^{+1.89}_{-1.71}$&6.7&8.97$^{+0.33}_{-0.3}$&0.85$^{+0.04}_{-0.03}$&26.207$^{+6.890}_{-6.890}$   \\
04483062+1623187&3855$\pm$6.77&60&8.709$^{+0.7509}_{-0.727}$&5.15&5.44$^{+0.24}_{-0.23}$&0.56$^{+0.02}_{-0.02}$&2.2890$^{+5.663}_{-5.663}$   \\
04484211+2106035&5215$\pm$14.51&60&74.9$^{+2.98}_{-4.82}$&5.97&8.71$^{+0.18}_{-0.28}$&0.85$^{+0.02}_{-0.03}$&2.3311$^{+3.602}_{-3.602}$   \\
04491296+2448103&4726$\pm$25.52&60&60.6$^{+2.37}_{-2.96}$&2.95&9.55$^{+0.21}_{-0.25}$&1.02$^{+0.02}_{-0.03}$&23.383$^{+5.086}_{-5.086}$   \\
04500069+1624436&4468$\pm$9.16&60&24.5$^{+1.38}_{-1.27}$&4.67&6.79$^{+0.19}_{-0.18}$&0.72$^{+0.02}_{-0.02}$&3.6727$^{+4.347}_{-4.347}$   \\
04510241+1458167&3945$\pm$8.25&60&16.3$^{+0.655}_{-0.617}$&2.0&7.1$^{+0.15}_{-0.14}$&0.62$^{+0.02}_{-0.02}$&5.2541$^{+4.629}_{-4.629}$   \\
04522352+1859489&4710$\pm$12.61&60&30.4$^{+1.2200}_{-1.150}$&2.89&6.81$^{+0.14}_{-0.13}$&0.74$^{+0.02}_{-0.01}$&1.7520$^{+3.751}_{-3.751}$   \\
04522385+1043099&3780$\pm$8.15&60&7.040$^{+0.368}_{-0.406}$&2.92&5.08$^{+0.13}_{-0.15}$&0.57$^{+0.02}_{-0.02}$&5.0041$^{+4.684}_{-4.684}$   \\
05054038+0627545&4859$\pm$14.25&60&41.7$^{+0.4100}_{-0.404}$&0.26&7.49$^{+0.06}_{-0.06}$&1.06$^{+0.02}_{-0.02}$&22.910$^{+4.659}_{-4.659}$   \\
05110971+1548574&3809$\pm$6.58&60&13.69$^{+0.3629}_{-0.349}$&1.35&6.99$^{+0.1}_{-0.09}$&0.86$^{+0.04}_{-0.04}$&30.224$^{+7.385}_{-7.385}$   \\ 
\enddata
\end{deluxetable*}

\end{document}